\documentclass[%
 reprint,
superscriptaddress,
 amsmath,amssymb,
 aps
]{revtex4-1}

\usepackage{graphicx}
\usepackage{dcolumn}
\usepackage{bm}
\usepackage[colorlinks]{hyperref}
\usepackage{subfigure}
\usepackage[final]{changes}
\usepackage[section]{placeins}

\newcommand{\fixme}[1]%
   {\begingroup{\color{blue}[NOTE: \textit{#1}]}\endgroup}

\graphicspath{{Figures/},
    {Figures/two-qubit/optimized/high/},
    {Figures/two-qubit/optimized/med/},
    {Figures/two-qubit/optimized/low/},
}

\begin{document}

\preprint{APS/123-QED}

\title{Validity of Born--Markov master equations for single and two-qubit systems} 

\author{Vasilii Vadimov}
\affiliation{QCD Labs, QTF Centre of Excellence, Department of Applied Physics, Aalto University, P.O. Box 13500,FI-00076 Aalto, Espoo, Finland}
\affiliation{MSP Group, QTF Centre of Excellence, Department of Applied Physics, Aalto University, P.O. Box 11000, FI-00076 Aalto, Espoo, Finland}
\affiliation{Institute for Physics of Microstructures, Russian Academy of Sciences, 603950 Nizhny Novgorod, GSP-105, Russia}
\author{Jani Tuorila}
\affiliation{QCD Labs, QTF Centre of Excellence, Department of Applied Physics, Aalto University, P.O. Box 13500,FI-00076 Aalto, Espoo, Finland}
\affiliation{MSP Group, QTF Centre of Excellence, Department of Applied Physics, Aalto University, P.O. Box 11000, FI-00076 Aalto, Espoo, Finland}
\affiliation{IQM, Keilaranta 19, FI-02150 Espoo, Finland}
\author{Tuure Orell}
\affiliation{Nano and Molecular Materials Research Unit, University of Oulu, P.O. Box 3000, FI-90014, Finland}
\author{J\"urgen Stockburger}
\affiliation{Institute for Complex Quantum Systems and IQST, University of Ulm, 89069 Ulm, Germany}
\author{Tapio Ala-Nissila}
\affiliation{MSP Group, QTF Centre of Excellence, Department of Applied Physics, Aalto University, P.O. Box 11000, FI-00076 Aalto, Espoo, Finland}
\affiliation{Interdisciplinary Centre for Mathematical Modelling, Department of Mathematical Sciences, Loughborough University, Loughborough, Leicestershire LE11 3TU, UK}
\author{Joachim Ankerhold}
\affiliation{Institute for Complex Quantum Systems and IQST, University of Ulm, 89069 Ulm, Germany}
\author{Mikko M\"ott\"onen}
\affiliation{QCD Labs, QTF Centre of Excellence, Department of Applied Physics, Aalto University, P.O. Box 13500,FI-00076 Aalto, Espoo, Finland}
\affiliation{VTT Technical Research Centre of Finland Ltd., QTF Center of Excellence, P.O. Box 1000, FI-02044 VTT, Finland}

\date{\today}

\begin{abstract}
The urgent need  for reliable simulation tools  to match the extreme accuracy needed to control tailored quantum devices highlights the importance of understanding open quantum systems and their modeling. 
To this end, we compare here the commonly used Redfield and Lindblad master equations against numerically exact results in the case of one and two resonant qubits transversely coupled at a single point to a Drude-cut ohmic bath.
All the relevant parameters are varied over a broad range which allows us to give detailed predictions about the validity and physically meaningful applicability of the weak-coupling approaches.
We characterize the accuracy of the approximate approaches by comparing the maximum difference of their system evolution superoperators with numerically exact results. After optimizing the parameters of the approximate models to minimize the difference, we also explore if and to what extent the weak-coupling equations can be applied at least as phenomenological models. Optimization may lead to an accurate reproduction of experimental data, but yet our results are important to estimate the reliability of the extracted parameter values such as the bath temperature.  
Our findings set general guidelines for the range of validity of the usual Born--Markov master equations and indicate that they fail to accurately describe the physics in surprisingly broad range of parameters, in particular at low temperatures. Since quantum-technological devices operate there their accurate modeling calls for a careful choice of methods.
\end{abstract}

\pacs{Valid PACS appear here}
\maketitle

\section{General introduction}
Precision control and measurement of quantum systems and devices~\cite{divincenzo2000,nielsen2000} have undergone great progress during the recent years. Important applications of this research field in both quantum computing~\cite{arute2019,kjaergaard2020superconducting,blais2020} and quantum heat engines~\cite{kosloff2014quantum,newman17,klatzow2019,lindenfels2019} call for in-depth studies of the accuracy of the corresponding theoretical models, especially for open quantum systems, where typically many approximations are utilized to render the problem computationally solvable~\cite{weiss2012,legett1987,breuerpetruccione02,rivas2011,ankerhold2001,braun2001,Salmilehto2014,deVega2017}. Most typically, Markovian master equations (MEs) are used, having been historically proven simple and effective tools in many scenarios where open quantum systems appear such as in the field of quantum optics~\cite{gardiner2004quantum}. The ME approach can be rigorously justified for the limit of weak coupling and a suitable separation of timescales between the system dynamics and the correlation time of the dissipative reservoir, which, in turn, depends heavily on the reservoir temperature. The most commonly used ME approaches include the Redfield and Lindblad equations~\cite{redfield1965, lindblad1976, gorini1976}. The latter is obtained from the former by an additional secular approximation to neglect rapidly oscillating terms in the density operator. 

However, such models of open quantum systems require a critical inspection in several important cases of contemporary science and technology. In quantum information, dissipation is typically weak for unitary gate operations but very high fidelities are pursued, thus setting stringent requirements for the accuracy of the theoretical models~\cite{tuorila2019}. For qubit reset in contrast~\cite{jones2013, tuorila2017, tuorila2019, valenzuela2006, grajcar2008, geerlings2013, jin2015}, temporarily strong dissipation is required at least effectively, possibly leading to non-trivial system--reservoir correlations~\cite{ankerhold2014,tuorila2019}, and consequently potential initialization errors.  In quantum thermodynamics, stronger reservoir coupling combined with finite-time operation gives typically rise to higher performance in terms of total power, and hence the parameter regimes of interest may be very different from those in quantum optics rendering many previous experimental verifications of the models inapplicable \cite{motz2018,wiedmann2020,senior2020}. 
Importantly, it is known that the consistency between some widely adopted models and fundamental physical concepts contradict each other, e.g., the stationary states contradict thermodynamical principles~\cite{levy2014,gonzalez2017,chiara2018}.

Historically, systematic experimental studies on the validity of the weak-coupling approaches have been challenged by the lack of systematic and predictable tunability of the relevant parameters such as the coupling strength to a broadband reservoir. The change of the coupling strength may also require to adjust the qubit frequency, which in turn may shift the qubit close to a spurious reservoir resonance in an unpredictable manner. Driving certain transitions may effectively provide a tunable dissipation, but renders the system intractable for the standard approaches assuming a non-driven system. Redesign and fabrication of a new sample may readily produce parameters in the desired range, but this is a very slow and resource-intensive approach. 

The recent development of a quantum-circuit refrigerator (QCR)~\cite{tan2017,silveri2017,sevriuk2019} has introduced the solid-state-qubit community with a simple device that provides orders of magnitude tunability in the system-reservoir coupling strength with minimal effect on the system parameters. Thanks to this tunability, the QCR has thus far been used to observe the Lamb shift arising from a broadband reservoir of an engineered quantum system~\cite{silveri2019}, and has the potential to enhance, for example, qubit initialization~\cite{Magnard2018}, quantum-thermodynamic devices~\cite{Pekola2015}, quantum-state-engineering protocols~\cite{Makhlin2001,Verstraete2009,Kastoryano2011,Rao2014}, and synthetic quantum matter~\cite{Houck2012,Fitzpatrick2017,Ma2019}. Together with the generally expanding experimental toolbox for quantum technology, QCR motivates us to benchmark the validity and accuracy of widely used approximate methods for open quantum systems against numerically exact solutions. Our theoretical study may thus work as a roadmap for various future experiments in the pursuit for computationally feasible and accurate models. 

In parallel to these developments, advanced descriptions of reduced open quantum dynamics have been formulated and applied to a variety of systems. These approaches are based on a non-perturbative representation of the reduced density matrix in terms of path integrals pioneered by Feynman and Vernon \cite{feynman1963, weiss2012}. Accordingly, path integral Monte Carlo techniques have been shown to provide insight into subtle qubit-reservoir correlations in regimes not accessible by other means \cite{kast2013}. Often more efficient and with a broader range of applicability are stochastic representations of the path integral dynamics \cite{strunz1997,stockburger2002}, in particular the Stochastic Liouville-von Neumann Equation (SLN) \cite{stockburger2004,schmidt2011,schmidt2013,wiedmann2020} and its version for ohmic dissipation (SLED, Stochastic Liouville Equation with Dissipation) \cite{stockburger1998,stockburger1999}.

This paper is organized as follows: After this general introduction to the field of research, we proceed in Sec.~\ref{sec:intro_models} to discuss in an introductory manner the different master equations used in our study and we especially elaborate on the Born--Markov approximation. In Sec.~\ref{sec:theory}, we introduce the microscopic Hamiltonian and define the error functional we use to study the difference between the evolution operators given by the different approaches. Sections~\ref{sec:single-qubit} and~\ref{sec:two-qubits} provide our most important numerical results on the single and two-qubit cases, respectively. 
Appendices~\ref{sec:born-markov} and~\ref{sec:stochastic-liouville} provide mathematical details of the used master equations and the numerically exact stochastic method. 

\section{Introduction to the models used}\label{sec:intro_models}
The formal requirements needed to achieve consistency between the Lindblad approach and the corresponding full microscopic model have been thoroughly studied~\cite{davie74,breuerpetruccione02}. 
Importantly, the dissipator terms in both Redfield and Lindblad equations do not directly correspond to any Hamiltonian operator of the microscopic model. 
They are rather a compact and approximate representation of the processes which amount to lowest-order emission and absorption of energy quanta. 

The Born--Markov (BM) approximation is at the heart of the Redfield equation. Here, one applies the lowest-order non-trivial perturbation theory for the system--reservoir coupling where the effect of the system--reservoir correlations on the evolution are neglected. In addition, one effectively applies coarse graining over time scales much longer than the characteristic time scale of the system Hamiltonian 
and assumes that the correlation time of the reservoir is much shorter than the resulting decay time. 
In addition to the above BM approximation, an additional assumption of a separable coupling forms the basis of the standard Redfield master equation. 

For a broadband reservoir, the correlation time is of the order of the thermal time $\hbar\beta=\hbar/(k_{\rm B}T)$, where $\hbar$ is the reduced Planck constant, $k_{\rm B}$ is the Boltzmann constant, and $T$ is the reservoir temperature. Thus, the BM approximation does not necessarily imply a white-noise limit. In fact, the separation of timescales characterizing the BM approximation is typically considered between the reservoir correlation time and the timescales of relaxation and dephasing processes caused by the system--reservoir interaction.

The reduced dynamics induced by the Redfield equation lacks a fundamental property of quantum channels: it is not completely positive. Even negative eigenvalues of the reduced density operator itself may appear. Neglecting quickly oscillating components of the density operator, a method generally referred to as the secular approximation, remedies this shortcoming and leads to the Lindblad equation. However, this advantage comes at the price of an additional condition of validity, namely, the level spacings of the system must greatly exceed the decay rates. This is typically a stricter requirement than the weak-coupling assumption in the BM approximation, and consequently the Redfield equation may in many cases provide a more accurate model although not guaranteeing complete positivity. We note that recently this shortcoming has been remedied by the derivation of a Lindblad-like master equation based on expansion in terms of the correlation between the bath and the system instead of the coupling strength \cite{alipour2020correlation}.

The secular approximation discussed above also includes a subtlety sometimes overlooked in the literature. Namely, the basis in which the secular approximation is carried out defines also the basis in which the dissipative transitions take place. This effect is pronounced in driven or multipartite systems where one may, for example, choose a local approach where the basis is chosen as the instantaneous eigenbasis of the individual constituents of the multipartite system or the global approach where one uses the eigenbasis of the full multipartite Hamiltonian taking into account its possible temporal trajectory~\cite{verso2010,gramich2014,levy2014,hofer2017,gonzalez2017,chiara2018,cattaneo2019}. 
Temporally local approach in the case of external driving has been observed to lead to unphysical results, for example, in Cooper pair pumping~\cite{Mottonen2008,Pekola2010,Salmilehto2011,Salmilehto2012}, and consitutes an interesting research direction. In this paper however, we focus  on non-driven systems and benchmark the validity of the local and global Lindbald equations and the Redfield equation against numerically exact qubit dynamics.

Exact methods beyond the BM family of open-quantum-system approaches are well established~\cite{feynman1963, grabert1988,stockburger1998,stockburger1999,stockburger2002,weiss2012}, but more complicated and computationally expensive. Beyond the Born approximation, reservoirs which are Gaussian can still be fully characterized by a two-time correlator. A somewhat loose but intuitively appealing characterization of this generalization can be given as follows: Gaussianity beyond the Born approximation implies that the high-order emission and absorption terms become relevant, but are reducible in the spirit of Wick's theorem. This scenario enables an exact description of the corresponding quantum dynamics through path integrals in the form of the Feynman--Vernon influence functional~\cite{feynman1963, grabert1988, weiss2012}. The influence functional is a non-local functional of the paths describing the propagation of the reduced density operator, and hence challenging to solve numerically. This inconvenience can be circumvented by a stochastic unraveling of the influence functional~\cite{stockburger2002} in a similar fashion to the Hubbard--Stratonovich transform. Thus one obtains a time-local master equation for the reduced system density operator which is otherwise of an equal computational complexity as the weak-coupling equations except that it is subject to two noise terms in the general case, or a single noise term in the case of ohmic dissipation~\cite{stockburger1998,stockburger1999}. We use the latter approach, referred to as stochastic Liouville equation with dissipation (SLED), as a benchmark to check the validity of the above-discussed BM approaches. The SLED has been proven to combine high numerical efficiency with high accuracy in broad ranges of parameter space together with the versatility to be adapted easily to various systems, see e.g. Refs. \onlinecite{schmidt2011,wiedmann2020}.
The existence of the noise terms necessitates one to ensemble average the density operators obtained for individual noise realizations to obtain the system density operator, which renders this method computationally much heavier than the weak-coupling approaches. Nevertheless, parallel computing can be utilized to obtain exact dynamics of low-dimensional open quantum systems.

\begin{figure}[ht!]
\includegraphics[width = \linewidth]{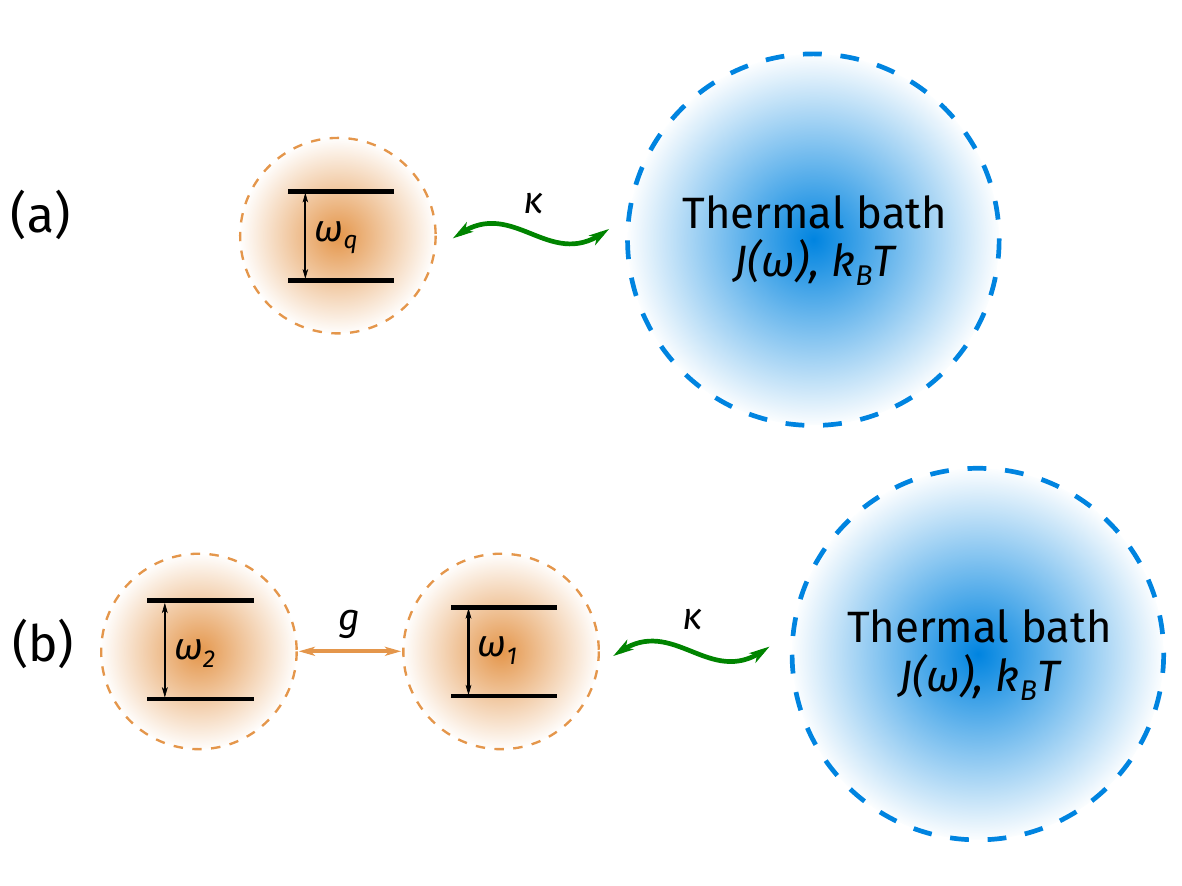}
\caption{(a) Single and (b) a two-qubit system coupled to the thermal bath, or reservoir, with the spectral density $J(\omega)$ and temperature $T$. The parameter $\kappa$ characterizes the coupling strength between the qubits and the reservoir, saturating to the decay rate in the zero-coupling limit. The angular frequencies of the qubits are denoted by $\{\omega_{k}\}_{k=\textrm{q},1,2}$. In the case of the two-qubit system, the qubits are coupled to each other with the coupling strength determined by a parameter $g$.}
\end{figure}

\section{Microscopic Hamiltonian and error of perturbative propagation}
\label{sec:theory}

In this paper, we study single and two-qubit systems embedded in a large number of reservoir degrees of freedom, a situation that generically appears in solid-state implementations. A typical realization of this scenario consists of electromagnetic modes interacting with a superconducting or a semiconductor qubit system, thus causing decoherence in the latter. If the quantum fluctuations caused by these reservoir modes are Gaussian in nature, they can be modeled by a set of harmonic oscillators bilinearly coupled to the qubit system. Accordingly, we assume a general Hamiltonian of the form
\begin{equation}
    \hat H = \hat H_{\rm S} + \sum\limits_k \hbar \Omega_k \hat b_k^\dag \hat b_k + \hat q \hat \xi,
    \label{eq:hamiltonian}
\end{equation}
where $\hat H_{\rm S}$ is the Hamiltonian of a single or two-qubit system, $\hat q$ is the system part of the system--reservoir coupling operator, and the corresponding reservoir operator is given by
\begin{equation}
    \hat \xi = \sum\limits_k g_k \left(\hat b_k^\dag + \hat b_k\right).
\end{equation}
For this type of a model, the effective impact of the reservoir onto the qubit system is characterized by the reservoir temperature $1/\beta$ and the weighted spectral density $J(\omega) = \pi \sum_k g_k^2 \delta(\omega - \Omega_k)/\hbar$. Below, we assume an Ohmic-type distribution with a high cutoff frequency $\omega_{\rm c}$ such that we obtain in the continuum limit
\begin{equation}
    J(\omega) = \frac{\eta \omega}{(1 + \omega^2 / \omega_{\rm c}^2)^2} .
    \label{eq:ohmic}
\end{equation}
The usual Drude cut-off term appears in squared form here to avoid divergences of the total noise power of the reservoir. Ohmic-type of reservoirs can be found in a broad class of qubit systems, particularly, in superconducting devices. In experiments, they may accurately capture qubit-reservoir interactions only in the moderate to high-frequency range, whereas at very low frequencies non-Ohmic behavior is typical and system dependent, for example in form of $1/f$ noise. Assuming a well calibrated system however, the latter are of minor relevance on the time scales of qubit control and error correction and are thus not studied in this work. Here, we rather present a detailed analysis of perturbative weak-coupling treatments in describing with sufficient accuracy the dissipative qubit dynamics in comparison to exact results.

 In order to quantify the difference in the numerical performance between the perturbative and the exact methods, we introduce the superoperator ${\mathcal T}(t)$ which transforms an initial reduced system density matrix $\rho_{\rm S}(0)$ to that at time $t$ as 
\begin{equation}\label{eq:T}
    \hat \rho_{\rm S}(t) = { \mathcal{ T}}(t)\hat\rho_{\rm S}(0).
\end{equation}
%
We estimate the accuracy of the BM approaches by calculating the distance
\begin{equation}
\Delta(t) = ||\overline{\mathcal T}_{\rm BM}(t)-\overline{\mathcal T}_{\rm SLED}(t)||/2,
\end{equation}
between normalized evolution superoperators of the BM type and of the corresponding numerically exact solution obtained with the SLED.
The normalization of a superoperator $\mathcal A$ is defined as $\overline{\mathcal A} = \mathcal A/||\mathcal A||$, where the Frobenius norm $||\cdot||$ is given by
\begin{equation}
    \replaced{||\mathcal A|| = \sqrt{\sum_{i=1}^{N^2}\sum_{j=1}^{N^2}|\mathcal{A}_{ij}|^2}}{\Delta(\hat A,\hat B) =\frac12\textrm{Tr}\left\{\sqrt{(\hat A- \hat B)^{\dag}[\hat A-\hat B]}\right\}}\ .
\end{equation}
with $\mathcal A_{ij}$ being the matrix elements of $\mathcal A$ in the chosen basis and $N$ denotes the dimension of the system Hilbert space, i.e., $N=2$ for the single-qubit system and $N=4$ for the two-qubit system.

In the case of time-independent Hamiltonians, the evolution superoperator corresponding to a BM master equation can be formally  represented as
\begin{equation}
    \mathcal T(t) = e^{\mathcal{L}t},
\end{equation}
where $\mathcal{L}$ is the Liouvillian superoperator of the open quantum system defined by $\dot\rho_{\rm S}(t)=\mathcal{L}\rho_{\rm S}(t)$ and including the non-unitary dissipative terms of the master equation. The non-unitary evolution superoperator $\mathcal T_{\mathrm{SLED}}(t)$ [see Eq.~(\ref{eq:T})] for the numerically exact SLED solution is constructed by solving the SLED using $N^2$ linearly independent initial conditions for the density operator $\hat \rho_{\rm S}(0)$.

Furthermore, a temporally independent figure to quantify the accuracy of a BM evolution is obtained by the maximum distance from the SLED defined as
\begin{equation}
    \Delta_\mathrm{max} = \max_{t\in[0,\infty)} \Delta(t)\, ,
\end{equation}
Since we are considering a non-driven decaying system, the maximum is attained in practice at a finite time. 

In a physical setup, the parameters which enter the above models are typically not be known a priori, but are adjusted after obtaining information from the system. Hence, we also carry out a study where we minimize $\Delta_{\rm max}$ by optimizing the parameters of the system, bath, and those determining their interaction. This optimization procedure may be interpreted as a simulation of a typical experimental situation in which the measurement data correspond to those of the SLED and the system parameters and dissipation rates are extracted to fit a BM model for the open quantum system (assuming structurally same system and system-reservoir coupling Hamiltonian). However, our detailed analysis demonstrates below that such a procedure may not always provide accurate predictions of the parameter values, especially in case of moderate or strong system--reservoir coupling, or if the interaction parameters with the reservoir are tuned during the evolution. One should be even more cautions about making extrapolations of experimental data deep into unmapped parameter regimes based on the fits.

\section{Single-qubit results}
\label{sec:single-qubit}

\begin{figure}[!ht]
\includegraphics[width=\linewidth]{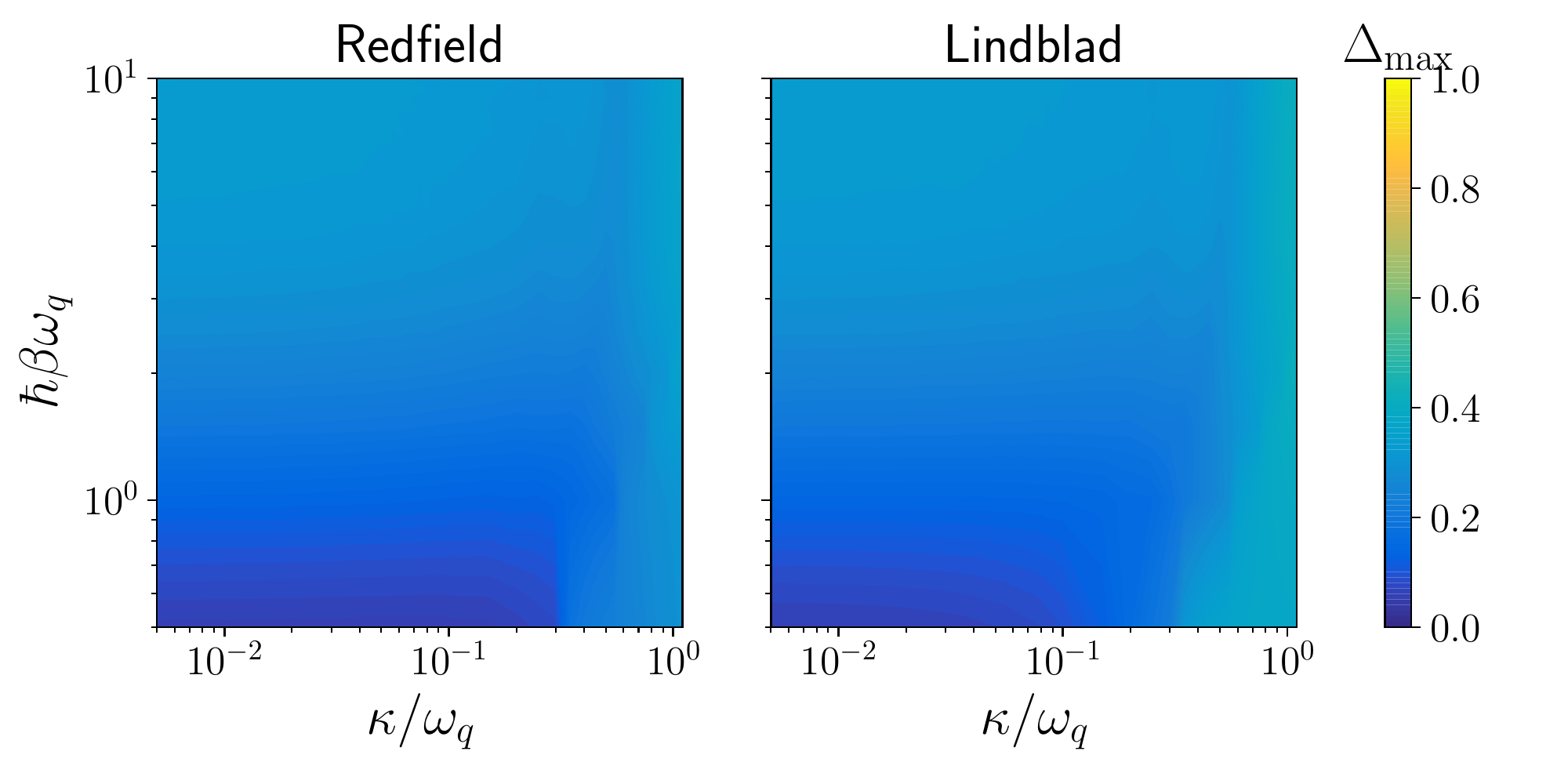}
\caption{Maximum value of the distance $\Delta_\mathrm{max}$ obtained with non-optimized values of $\kappa$, $\beta$, and the qubit angular frequency $\omega_{\mathrm q}$. The non-unitary temporal-evolution superoperator $\mathcal T_{\mathrm{SLED}}(t)$ [see Eq.~\eqref{eq:T}] is constructed numerically by solving the SLED using initial states $\hat \rho_\mathrm{SLED}^\mathit{x}(0) = |\sigma_{ x},+\rangle\langle \sigma_{x},+|$, $\hat \rho_\mathrm{SLED}^{y}(0) = |\sigma_\mathit{y},+\rangle\langle \sigma_\mathit{y},+|$, $\hat \rho_\mathrm{SLED}^{z}(0) = |\sigma_{z},+\rangle\langle \sigma_{z},+|$, and $\hat \rho_\mathrm{SLED}^{I}(0) = \frac12 \hat I$, where $|\sigma_{i},+\rangle$ is the excited eigenstate of the $\hat \sigma_{i}$ operator with $i =  x,y,z$, and $\hat I$ is the identity operator. The number of samples in the SLED solutions is $N_{\mathrm{traj}} = 10^5$ and we have used the cut-off frequency $\omega_{\rm c}/\omega_{\rm q} = 50$. For each pair $\{\kappa, \beta\}$, the trace distance is calculated for times $[0, 10\kappa_T^{-1}]$, where $\kappa_T = \kappa\coth(\hbar \beta \omega_{\rm q}/2)$.}\label{fig:maxorigDelta}
\end{figure}

\begin{figure}[!ht]
\includegraphics[width=\linewidth]{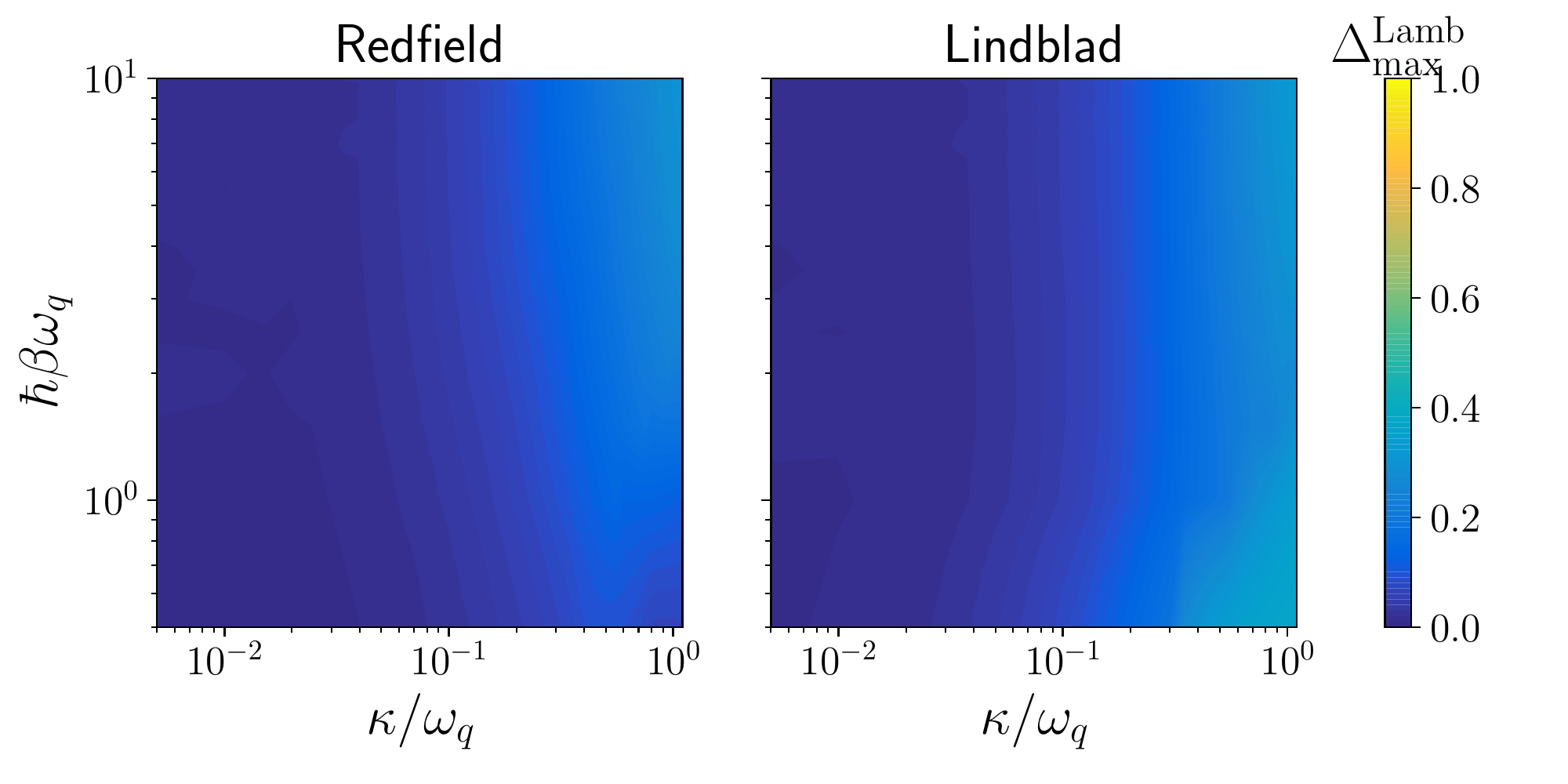}
\caption{Maximum value of the distance $\Delta_\mathrm{max}$ obtained with non-optimized values of $\kappa$, $\beta$, and the Lamb-shift-corrected qubit angular frequency $\Omega_{\rm q}$~(see Eq.~\eqref{eq:lamb}).}\label{fig:lamb-max-delta}
\end{figure}

%

\begin{figure}[ht!]
\includegraphics[width=\linewidth]{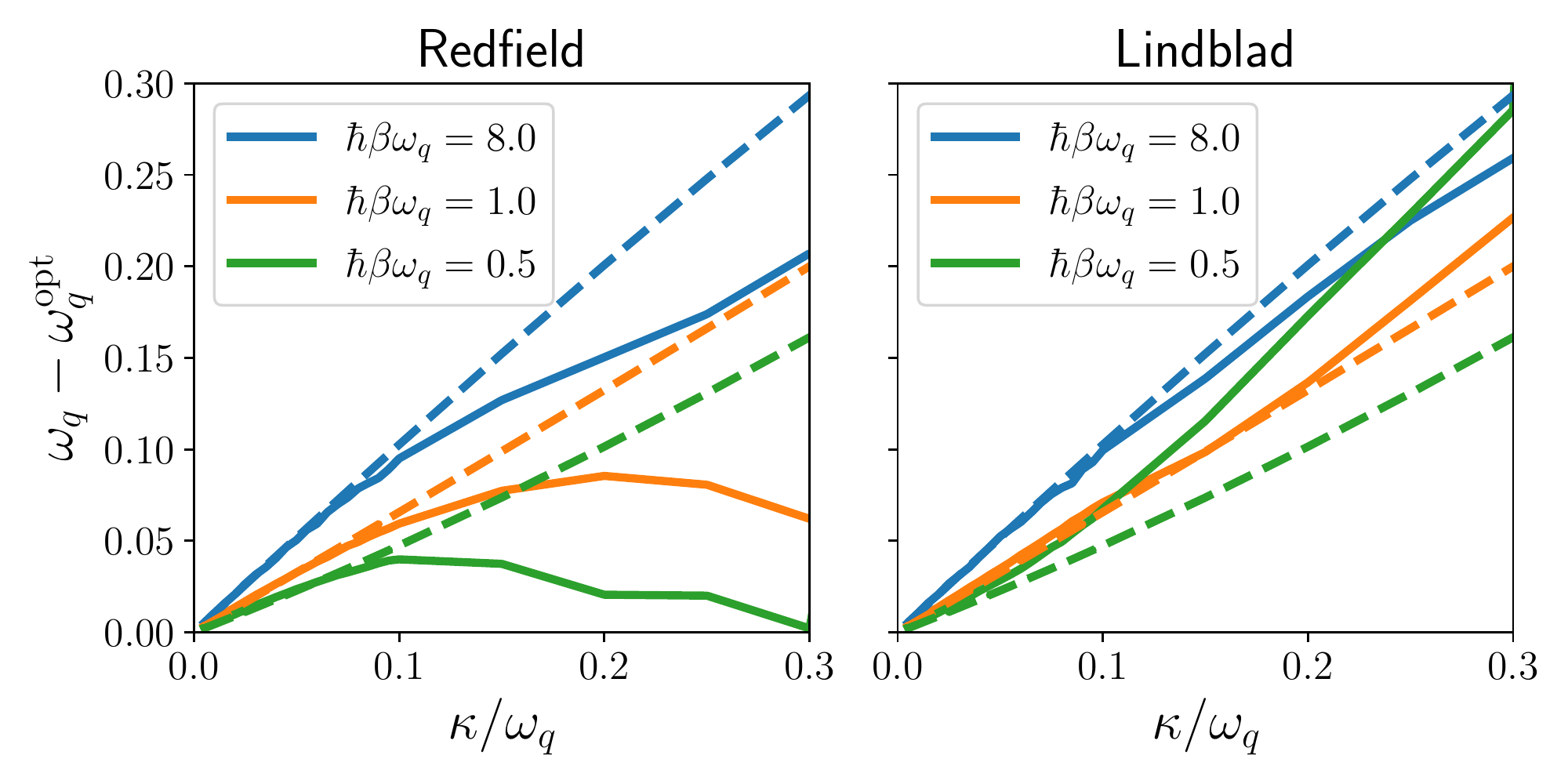}
\caption{Correction to the qubit frequency obtained by the optimization procedure $\omega_{\rm q} - \omega_{\rm q}^\mathrm{opt}$ (solid lines) and the analytically predicted Lamb shift $\omega_{\rm q} - \Omega_{\rm q}$ (dashed lines).
}
\label{fig:1qubit-lamb-shift}
\end{figure}

\begin{figure*}[ht!]
\includegraphics[width=\linewidth]{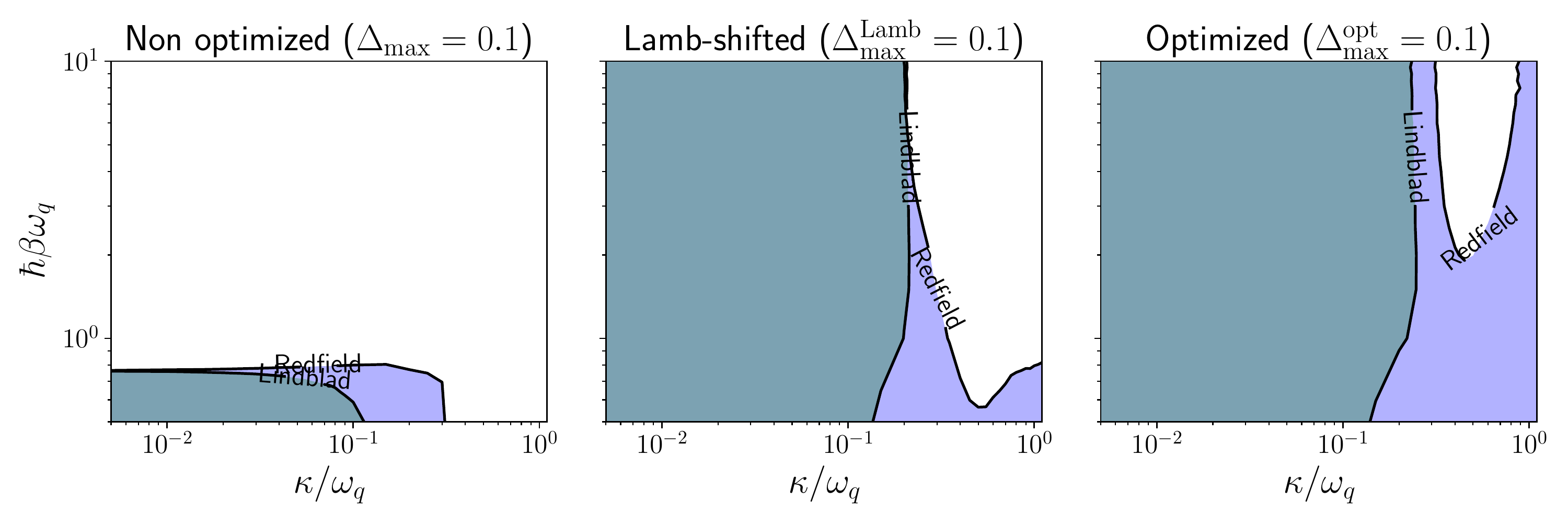}
\caption{A diagram showing the maximal error contour lines $\Delta_\mathrm{max} = 0.1$ in the case of a single qubir coupled to the environment. From left to right, we have the non-optimized (left), Lamb-shift-corrected (middle), and optimized (right) Lindblad (green) and Redfield (blue) data. 
}
\label{fig:1qubit-phase-diagram}
\end{figure*}

We start with the single qubit case where the system 
\begin{equation}
    \hat H_{\rm S} = \hbar \omega_{\rm q} \hat \sigma^+ \hat  \sigma^- ,
\end{equation}
is coupled to the bath through the operator $\hat q = \hat \sigma^+ + \hat \sigma^- = \hat \sigma^{x}$ with the parameter $\kappa = 2 \hbar \eta \omega_{\rm q}$ controlling the coupling strength to the bath. Here, $\hat{\sigma}^+=(\hat{\sigma}^-)^\dagger=|\textrm{e}\rangle\langle\textrm{g}|$, where $|\textrm{g}\rangle$ and $|\textrm{e}\rangle$ are the ground and the excited state of the qubit, respectively.

In Fig.~\ref{fig:maxorigDelta} we show the  distance $\Delta_\mathrm{max}$ for the  BM solutions (Redfield and Lindblad) calculated for dimensionless parameters $\kappa/\omega_{\rm q}$ and $\omega_{\rm q}\hbar\beta$. As expected their performance deteriorates with increasing qubit-reservoir coupling, where the Redfield covers a broader domain with acceptable accuracy.

What is not expected at the first glance is that the accuracy of BM approaches deteriorates with the decrease of the bath temperature while staying in the weak coupling regime $\kappa \ll \omega_{\rm q}$. The possible reason of this could be disregarding the Lamb shift which should be significant in the low temperature case. In order to verify this,  we perform BM calculations including the Lamb-shifted qubit frequency $\Omega_{\rm q}$ given by the following expression~\cite{weiss2012,tuorila2019}:
\begin{equation}
    \Omega_{\rm q} = \omega_\mathrm{eff}\left\{
    1 + 2 K \left[\mathop{\mathrm{Re}} \psi\left(i \frac{\hbar \beta \omega_\mathrm{eff}}{2\pi}\right) - \ln \left(\frac{\hbar \beta \omega_\mathrm{eff}}{2\pi}\right)\right]
    \right\}^{1/2},
    \label{eq:lamb}
\end{equation}
where $\psi(x)$ is the digamma function, $\omega_\mathrm{eff}=G(\omega_{\rm q} / \omega_{\rm c})^{K/(1-K)} \omega_{\rm q}$, $K = \kappa / (2\pi \omega_{\rm q})$, $G=[\Gamma(1-2K)\cos(\pi K)]^{1/[2(1-K)]}$, and $\Gamma(x)$ is the gamma function. Note that for simplicity, we do not differentiate here between the Lamb and Stark shifts, but refer to the total environment-induced frequency shift of the system as the Lamb shift. 

Figure \ref{fig:lamb-max-delta} shows the distance between the SLED and BM solutions with the account of a Lamb-shifted qubit frequency. The performance of the weak-coupling approaches is significantly improved and the regime of their applicability is extended to lower temperatures.

We also provide an optimized BM solution by finding for a given set of parameters those values for $\{\kappa^\mathrm{opt},\beta^\mathrm{opt},\omega_{\rm q}^\mathrm{opt}\}$  that minimize the maximal value of the distance $\Delta_\mathrm{max}$, i.e.  $\Delta_\mathrm{max}^\mathrm{opt} = \Delta_\mathrm{max}(\kappa^\mathrm{opt},\beta^\mathrm{opt},\omega^\mathrm{opt})$. Technically, the optimization is carried out using the Powell minimization method available as one of the standard methods in the SciPy numeric library~\cite{virtanen2020scipy}.

It is an interesting question whether or not the optimization procedure can capture the Lamb shift originating from the interaction with the environment. In order to answer this, we compare the correction to the qubit frequency obtained by the optimization procedure $\omega_{\rm q} - \omega_{\rm q}^\mathrm{opt}$ with the analytically predicted Lamb shift $\omega_{\rm q} - \Omega_{\rm q}$. This comparison is shown in Fig.~\ref{fig:1qubit-lamb-shift}. Apparently, the corrections are consistent with each other only in the weak coupling regime $\kappa / \omega_{\rm q} \lesssim 0.1$.


Figure~\ref{fig:1qubit-phase-diagram} displays areas of acceptable accuracy of non-optimized and optimized BM methods, where we consider a maximal distance of $\Delta_\mathrm{max}=0.1$ as a threshold.
While we confirm that the non-optimized BM solutions without the Lamb-shift  correction are limited by sufficiently elevated temperatures $\hbar \beta \omega_{\rm q} \lesssim 1$ and low coupling strengths between the bath and the qubit $\kappa / \omega_{\rm q} \ll 1$, we
 find that the optimized BM solutions as well as BM solutions with the Lamb-shift taken into account approximate the SLED solution quite well even at lower temperatures up to $\hbar \beta \omega_{\rm q} \approx 10$. This implies that the dynamics of the qubit can be effectively, i.e. by properly tuned parameters, captured  by Markovian dynamics in the weak coupling limit. However, in a broad range the values of these optimized parameters differ substantially from the bare ones (see Figs.~1--3 in the Supplemental Materials~\footnote{Supplemental material is available at (the link is to be added)}) 
 and even physically cannot always be considered as meaningful. In fact, they are either outside the range of formal validity of the underlying approximations of the BM approaches and/or are not reasonable given typical experimental set-ups. More specifically, the optimization parameters should be trusted only in the weak coupling regime $\kappa \lesssim 0.3 \omega_{\mathrm q}$.

\section{Two-qubit results}
\label{sec:two-qubits}

\begin{figure*}
    \centering
    \subfigure[High temperature $\hbar \omega_1 \beta = 0.1$]{
    \includegraphics[width=\linewidth]{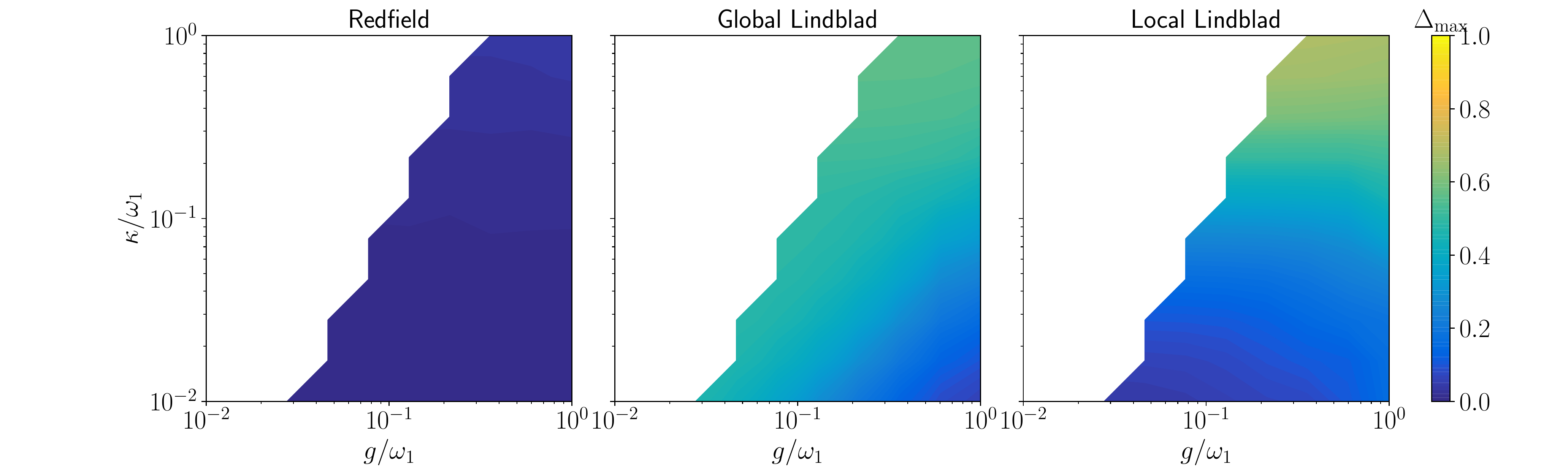}}
    \subfigure[Intermediate temperature $\hbar \omega_1 \beta = 1$]{
    \includegraphics[width=\linewidth]{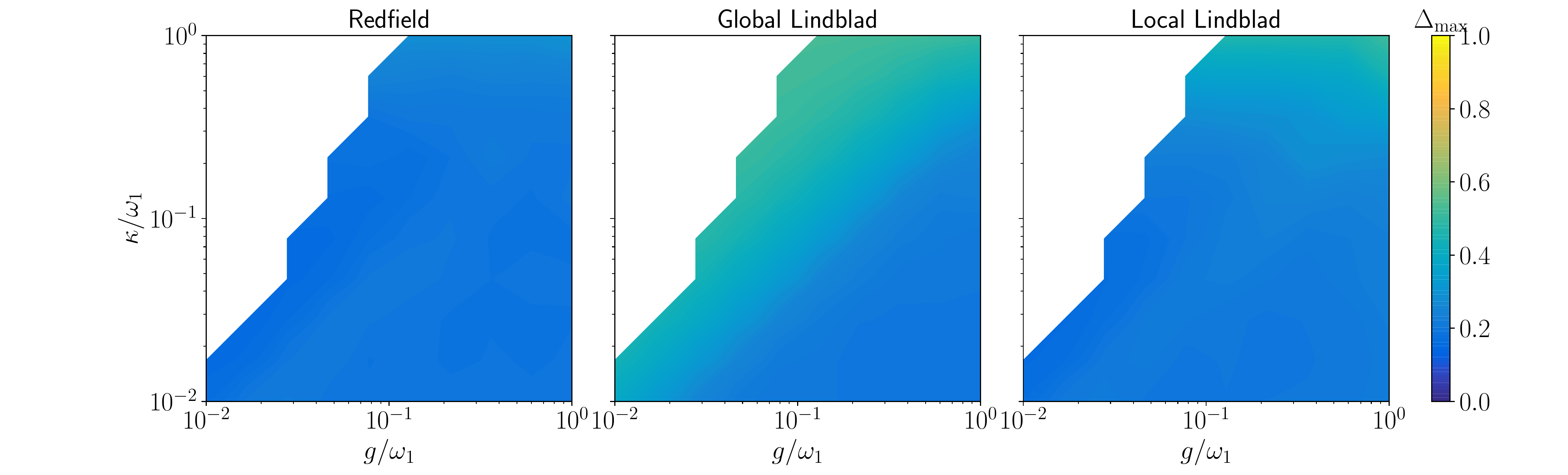}}
    \caption{Distance $\Delta_\mathrm{max}$ for non-optimized values of parameters at the (a) high temperature $\hbar \omega_1 \beta = 0.1$ and (b) intermediate temperature $\hbar \omega_1 \beta = 1$. 
    The number of samples in the SLED solutions is $N_{\mathrm{traj}} = 10^4$ and we have used the cut-off frequency $\omega_{\rm c}/\omega_{1} = 50$. For each pair $\{g, \kappa\}$, the distance is calculated for times $[0, 2 \times 10^3 \omega_1^{-1}]$
    The white color corresponds to the range of parameters for which the steady state has not been achieved.}
    \label{fig:2qubit-non-opt}
\end{figure*}

\begin{figure}
    \centering
    \includegraphics[width=\linewidth]{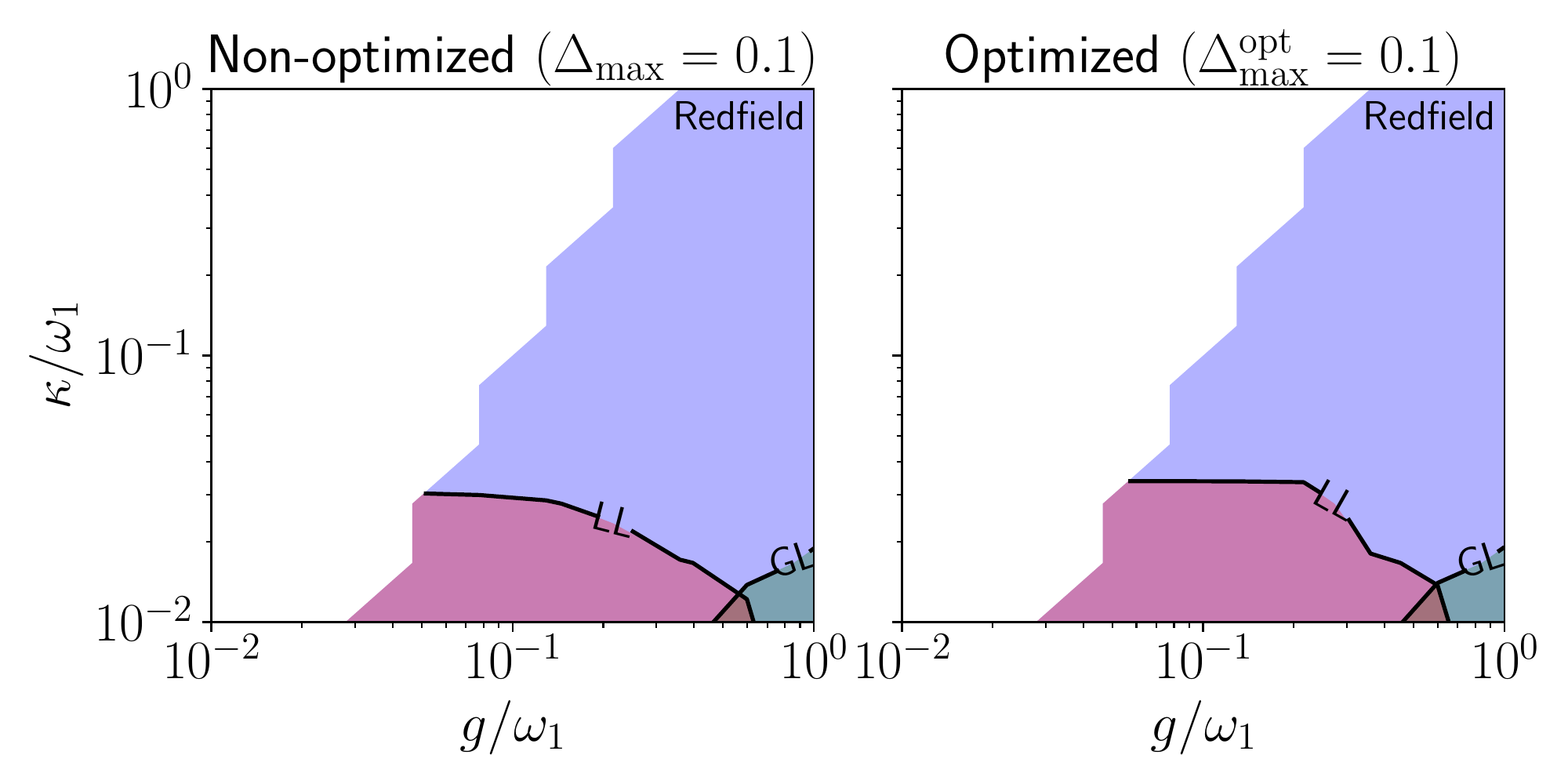}
    \caption{A diagram for the two-qubit case showing contour lines for $\Delta_{\rm max} = 0.1$ at high temperature $\hbar \beta \omega_1 = 0.1$. The data are for the non-optimized (left) and the optimized (right) Redfield (blue), GL (green) and LL (red) data.
    The white color corresponds to the range of parameters for which the steady state has not been achieved.}
    \label{fig:2qubit-diagram-high}
\end{figure}

\begin{figure*}
    \centering
    \subfigure[High temperature $\hbar \omega_1 \beta = 0.1$]{
    \includegraphics[width=\linewidth]{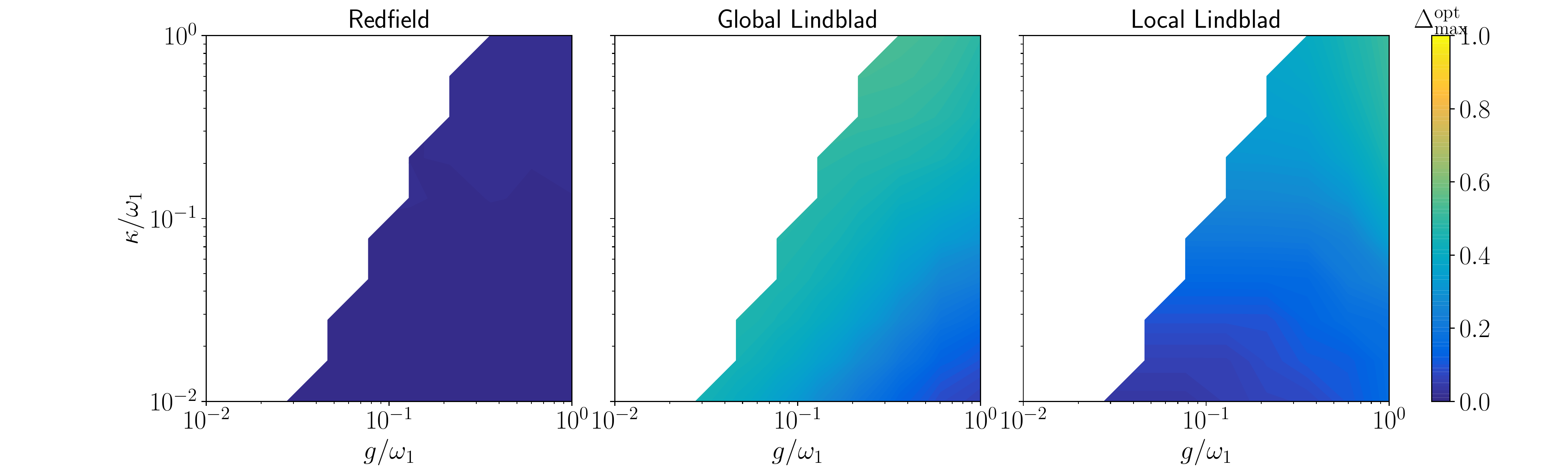}}
    \subfigure[Intermediate temperature $\hbar \omega_1 \beta = 1$]{
    \includegraphics[width=\linewidth]{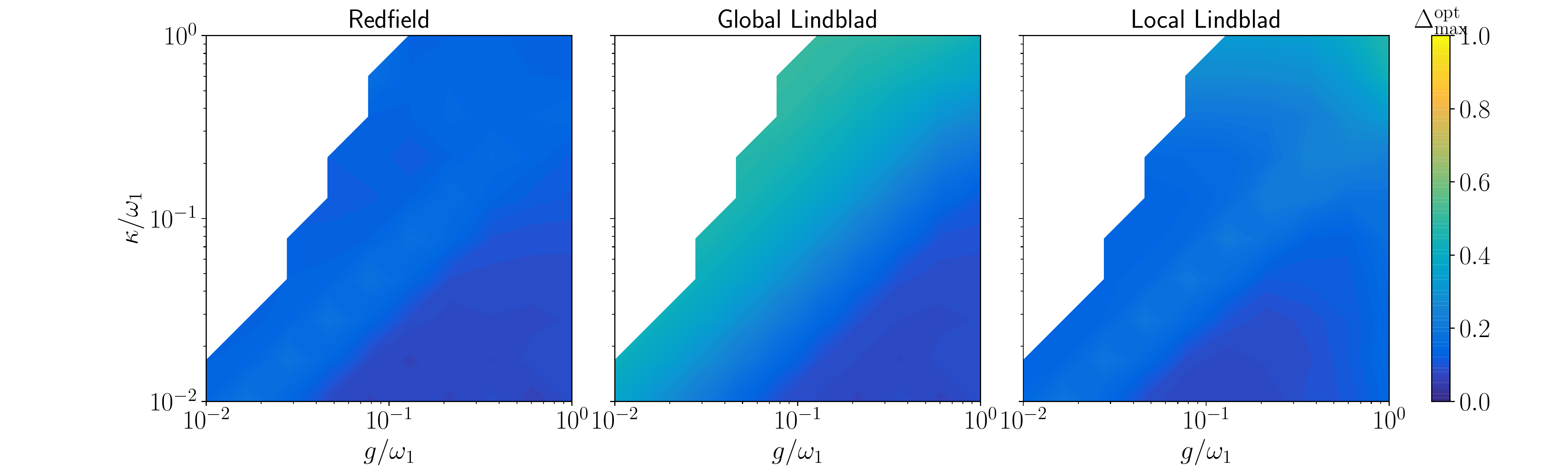}}
    
    \caption{Distance $\Delta_\mathrm{max}^\mathrm{opt}$ for optimized values of parameters at the (a) high temperature $\hbar \omega_1 \beta = 0.1$ and (b) intermediate temperature $\hbar \omega_1 \beta = 1$. The white color corresponds to the range of parameters for which the steady state has not been achieved.
    }
    \label{fig:2qubit-opt}
\end{figure*}

\begin{figure*}
    \includegraphics[width=\linewidth]{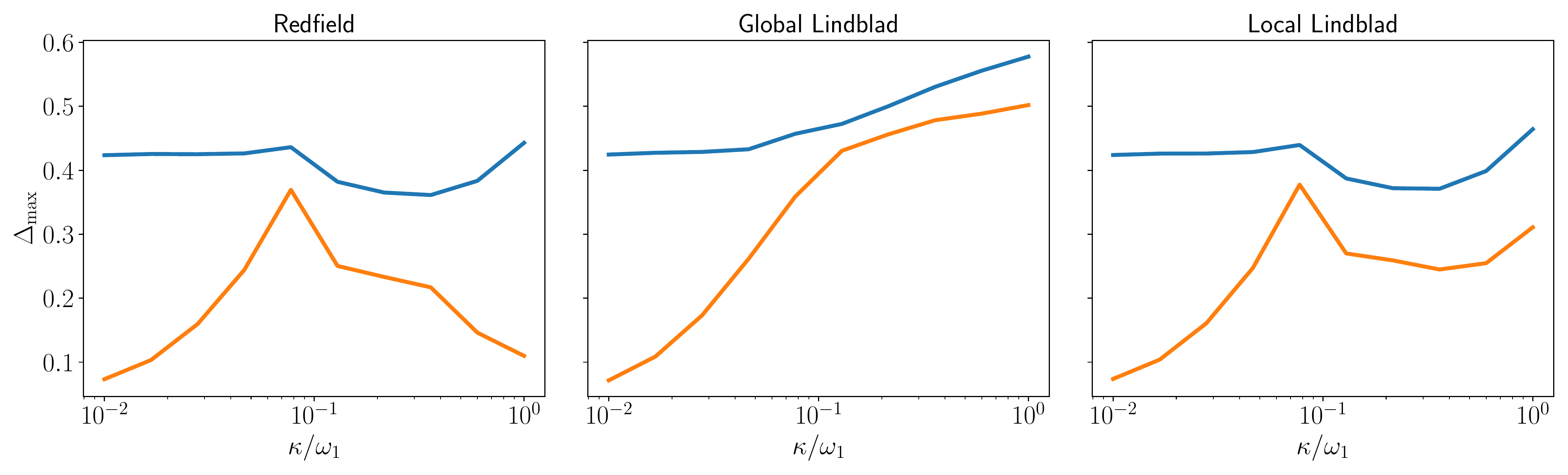}
    \caption{Distance $\Delta_\mathrm{max}$ for non-optimized (blue) and optimized (orange) values of parameters at the low temperature $\hbar \omega_1 \beta = 5$, $g = 0.1 \omega_1$. The number of samples in the SLED solutions is $N_{\mathrm{traj}} = 5 \cdot 10^5$ and we have used the cut-off frequency $\omega_{\rm c}/\omega_{1} = 50$. For each value of $\kappa$, the distance is calculated for times $[0, 10^3 \omega_1^{-1}]$.}
    \label{fig:2qubit-low}
\end{figure*}

Let us consider here two linearly coupled qubits described by the system Hamiltonian  
\begin{equation}
    \hat{H}_{\rm S} = \hbar \omega_1 \hat{\sigma}_1^+ \hat{\sigma}_1^- + \hbar \omega_2 \hat{\sigma}_2^+ \hat{\sigma}_2^- + \frac{\hbar g}{2} \hat{\sigma}_1^{x} \hat{\sigma}_2^{x},
    \label{eq:2qubit-hamiltonian}
\end{equation}
where the angular frequency of qubit $k$ is denoted by $\omega_k$ and the qubit--qubit coupling strength is denoted by $g$. In the total Hamiltonian, only the first qubit denoted by the subscript 1 is coupled to the bath through the operator $\hat q = \hat \sigma_1^x$. We consider two types of weak-coupling treatments: For weak qubit--qubit coupling, one typically uses the local Lindblad (LL) master equation, where the dissipators induce transitions between the eigenstates of the bare qubit 1. In contrast, the global Lindblad (GL) describes transitions in the two-qubit eigenbasis. 

For LL, the master equation of the reduced density operator of the two-qubit system is expressed as
\begin{multline}
    \frac{\textrm{d}\hat \rho_{\rm S}}{\textrm{d}t} = -\frac{i}{\hbar} \left[\hat H_{\rm S}, \hat \rho_{\rm S}\right] +\\ \frac{\kappa}{2} \left[N(\omega_1) + 1\right] \left[ 2 \hat \sigma_1^- \hat \rho_{\rm S} \hat \sigma_1^+ - \left\{\hat \sigma_1^+ \hat \sigma_1^-, \hat \rho_{\rm S}\right\}\right] + \\
    \frac{\kappa}{2} N(\omega_1)\left[2 \hat \sigma_1^+\hat \rho_{\rm S} \hat \sigma_1^- - \left\{\hat \sigma_1^- \hat \sigma_1^+, \hat \rho_S\right\}\right],
\end{multline}
where  $N(\omega) = 1 / [\exp(\hbar \beta \omega) - 1]$ is the bosonic occupation. 

Further analytical progress is possible if one applies the rotating-wave approximation, i.e., replaces the qubit--qubit coupling term $\sigma_1^x \sigma_2^x$ in the Hamiltonian by $\hbar(\hat \sigma_1^+ \hat \sigma_2^- + \hat \sigma_1^- \hat \sigma_2^+)  g / 2 $. Consequently, we arrive at the following equation of motion which is expressed, for simplicity, in the zero-temperature limit:
\begin{equation}
    \frac{\textrm{d}^2\langle\hat \sigma_2^-\rangle}{\textrm{d}t^2} + \left(i\delta_{12} + \frac{\kappa}{2}\right) \frac{\textrm{d}\langle \hat \sigma_2^-\rangle}{\textrm{d}t} + \frac{g^2}{4} \langle \hat \sigma_2^- \rangle = 0,
\end{equation}
where $\delta_{1,2} = \omega_1 - \omega_2$ is the detuning between the qubits. Thus, the exponential relaxation of the system is characterized by two complex-valued decay rates $\lambda_{1}$ and $\lambda_2$ 
which in case of resonant qubits $\delta_{1,2} = 0$ turn out to be
\begin{equation}
    \lambda_{1,2} = \frac{\kappa}{4} \pm \frac{1}{2} \sqrt{\frac{\kappa^2}{4} - g^2} .
\end{equation}
Accordingly, in the case of weak damping, $\kappa \ll g$, qubit~2 displays underdamped oscillations towards its bare ground state with the amplitude relaxation rate $\kappa / 4$. In the opposite limit, $\kappa \gg g$, a separation of time scales occurs, where one of the resulting rates $\kappa / 2$ by far exceeds the other $g^2 / (2\kappa)$. The latter rate determines the full equilibration time scale of the system. From a simulation point of view, this phenomenon implies that the computation of the asymptotic long-time behavior requires significant computational resources. Hence, instead of simulating the full-length equilibration dynamics of the system, we monitor the quantum evolution only for a fixed time interval $[0, T]$, where we use $T = 2 \times 10^3/\omega_1$, and subsequently analyze whether the system converged into its steady state or not. To this end, we compute the least negative eigenvalue $\lambda_\mathrm{min}$ of the Liouvillian and compare $T$ with the relaxation time  estimated by $-3 / \mathop{\mathrm{Re}} (\lambda_\mathrm{min})$. If $T$ exceeds the relaxation time, we conclude that the system has reached equilibrium.

According to this procedure, we study the accuracy of the LL equation, the GL equation, and the Redfield equation. See Appendix~\ref{sec:born-markov} for details of the weak-coupling equations. For the sake of clarity, 
we focus on resonant qubits $\omega_1 = \omega_2$ throughout this section. In the opposite case, $|\omega_1 - \omega_2| \gg g$, the coupling between the qubits appears as a weak perturbation to the local eigenstates and it is expected that effect of the reservoir which directly interacts with the first qubit, may be described using the weak-coupling approaches for the second qubit even in the regime $\kappa\gtrsim g$, provided that $\kappa\ll\omega_1,\omega_2$. 


For which parameters does one expect that the weak-coupling approaches provide reliable predictions? In order to justify the secular approximation, the coupling strength to the reservoir characterized by $\kappa$ should be very weak compared with the smallest distance between the energy levels of the Hamiltonian~(\ref{eq:2qubit-hamiltonian}) which is equal to $g$ for the qubits in resonance. Thus for our case of resonant qubits, we expect the GL equation to be valid only for $\kappa \ll g$. 
The Redfield equation partially cures this deficiency since it relies only on the BM approximations but does not invoke the secular approximation. 
Strictly speaking, all three approaches call for $\hbar \kappa \beta \ll 1$ to justify the Markov approximation (see the Appendix~\ref{sec:born-markov} for the details).

Figure~\ref{fig:2qubit-non-opt} shows  the maximum distance  between the SLED solution and each weak-coupling approach, $\Delta_\mathrm{max}$, as a function of the qubit--bath and qubit--qubit coupling strengths, $\kappa$ and $g$ respectively, for high ($\hbar \omega_1 \beta = 0.1$) and intermediate ($\hbar \omega_1 \beta = 1$) temperatures. We observe that in the high-temperature regime the Redfield equation reproduces the exact dynamics for the whole considered range of parameters, whereas the Lindblad approaches based on secular approximations have significant limitations. As expected, the GL approach is valid only in the region $g \gg \kappa$, while the LL is the most accurate for $\kappa / \omega_1, g/\omega_1 \ll 1$.
The poor performance of the GL equation is due to the neglected slowly oscillating terms with the frequency of order $g$ owing to the secular approximation. These terms appear due to the splitting of the resonant qubit levels which implies that the resonant situation is the most problematic one, whereas for non-resonant qubit systems with all the levels sufficiently separated, the accuracy of the GL equation is expected to be higher.
Interestingly, the situation is different at intermediate temperatures $\hbar \beta \omega_1 = 1$ even for the Redfield treatment as shown in Fig.~\ref{fig:2qubit-non-opt}(b). The solutions given by the SLED and the BM approaches are noticeably different which is largely explained by the lack of the Lamb shift for the weak qubit--bath coupling and by the violation of the BM conditions for strong coupling. 

To illustrate the regimes of validity of the weak-coupling approaches, we show in Fig.~\ref{fig:2qubit-diagram-high}(a) the contours of the maximum deviation $\Delta_\textrm{max}=0.1$ in the two-dimensional parameter space considered. Note that in practice, the tolerable threshold for the deviation depends on the application of interest.

Here, we optimize the parameters of the two-qubit system and the bath used in the BM equations in order to minimize the distance $\Delta_\mathrm{max}$ with respect to the SLED solution. To account for the two-qubit Lamb shift, we minimize $\Delta_\mathrm{max}$ by adding Hamiltonian terms to the system Hamiltonian such that the new terms commute with the original Hamiltonian of the system. Provided that the spectrum of the Hamiltonian is non-degenerate, an arbitrary operator commuting with it can be expressed as
\begin{equation}
    \hat H'_\textrm{S} = \sum\limits_{j=0}^{N-1} a_j (\hat H_\textrm{S})^j,
\end{equation}
where $N=4$ for the two-qubit system and $\{a_j\}$ are free real-valued parameters. The coefficient $a_0$ only contributes to the unobservable global phase, and hence can be considered to be zero without loss of generality. Thus we optimize $\Delta_\mathrm{max}$ by adjusting the parameters $a_1$, $a_2$, $a_3$, $\kappa$ and $\beta$.
The optimized distance $\Delta_\mathrm{max}^\mathrm{opt}$ as a function of the coupling parameters $g$ and $\kappa$ is shown in Fig.~\ref{fig:2qubit-opt} for high and intermediate temperatures. The optimization clearly improves the accuracy of the BM approaches, especially in the weak coupling case. However, the qualitative behavior of the maximum distance as a function of the parameters remains unchanged in the optimization as also visible in the $\Delta_\mathrm{max}^\mathrm{opt}=0.1$ contours in Fig.~\ref{fig:2qubit-diagram-high}(b). 

Figure~\ref{fig:2qubit-low} provides $\Delta_\mathrm{max}$ at low temperature $\hbar \beta \omega_1 = 5$. Owing to the large number of samples needed at this temperature to reach numerical convergence for SLED, we have chosen to fix  $g = 0.1\times\omega_1$. With these parameters, the non-optimized weak-coupling approaches fail to describe the dynamics of the system even for $\kappa\ll g\ll\omega_1$, whereas the optimized solutions display much better performance, at least in certain parameter ranges, thanks to the proper account of the Lamb shift caused by the environment. 
Indeed, we also observe that even the optimized approaches yield high error near $g \approx \kappa$, i.e., close to the point of critically damped dynamics for qubit 2. Interestingly, we find that the maximum error actually decreases for the Redfield and local Lindblad equations for couplings stronger than the critical point, at least up to about $\kappa=\omega_1$. 

Although the optimized weak-coupling model may relatively accurately yield the exact dynamics it may still be, depending on the parameters, that it misses important physics which may render the extracted parameter values questionable. Fortunately, this is not the case in our study except for the highest considered qubit--bath coupling strengths as we further discuss in the Supplemental Materials~\cite{Note1}.

\section{Summary and conclusions}

Let us summarize our main results. For a single qubit with angular frequency $\omega_{\rm q}$ and realistic  microscopically derived model parameters, we find that the approximate approaches are valid ($\Delta_\mathrm{max}<0.1$) only at high bath temperatures,   $T\gtrsim 1.5\times\hbar\omega_{\rm q}/k_{\rm B}$ ($\omega_{\rm q}\hbar\beta\lesssim 0.7$), and for low relaxation rates, $\kappa<0.1\times \omega_{\rm q}$, for Lindblad and slightly greater values $\kappa<0.3\times\omega_{\rm q}$ for Redfield. The optimization of the system parameters allows to expand the validity of these approaches to lower temperatures, at least down to $T \approx 0.1\times\hbar \omega_1 / k_{\rm B}$ ($\omega_{\rm q}\hbar\beta\approx 10$), in the weak-coupling regime $\kappa \ll \omega_{\rm q}$. 

Note that whereas the non-optimized results provide us information on the implications of the approximations carried out to arrive from the microscopic model to the approximate master equations, the optimized results may be considered as a test of the accuracy of the approaches as phenomenological models. In typical experiments, the latter case is important since the microscopic parameters may be inaccessible. 

In the two-qubit case and at high bath temperatures, $T\gtrsim 10\omega_1/k_{\rm B}$ ($\omega_1\hbar\beta\lesssim 0.1$), we find that the Redfield equation is valid ($\Delta_\mathrm{max}<0.1$) in the whole parameter range studied ($\omega_1/100<\kappa,g<\omega_1$ and $\kappa\lesssim  10g^2/\omega_1$). Interestingly, the global Lindblad approach is valid only for large enough qubit--qubit coupling $g\gtrsim 40\kappa$, whereas the local approach is valid only for $\kappa\lesssim 0.03\times\omega_1$ and $g\lesssim 0.6\times\omega_1$. 
Optimization of the model parameters extends validity of the local Lindblad method to intermediate qubit--qubit coupling $g \lesssim \omega_1$ and does not significantly change the validity bounds of the other approximate approaches.
At intermediate and low $T$, we find $\Delta_\textrm{max}>0.1$ essentially in the whole parameter range considered.

With decreasing temperature, we observe a dramatically increasing deviation between the non-optimized approximate and the exact dynamics. Optimization 
cures this discrepancy at weak coupling, but peculiarly, the point $\kappa\approx g$ seems problematic for weak-coupling approaches. We attribute this behavior to the failure of the BM equations to correctly capture the effect of the environment to the modes of the system in this point of critical damping for the qubit with an indirect coupling to the bath through the other qubit. 

We conclude that as expected, the non-optimized BM equations provide accurate dynamics only for weak coupling and high enough temperature for both single and two-qubit systems. This in turn excludes them as sufficiently reliable tools for many important experimental scenarios. For superconducting qubits, we may have, for example, $\omega_{\rm q} \approx 2\pi\times10$~GHz and $T\approx 40$ mK, and hence $\omega_{\rm q}\hbar\beta\approx 10$. 

Using the optimization procedure we demonstrated that the dynamics of the systems can be described by BM equations in broader ranges of parameters. 
However, the fitting experimental data with BM equations beyond their regimes of validity may, in some cases, yield physically misleading parameter values for the system, the bath, and their coupling strengths. 




\acknowledgments

This research  was  financially  supported  by  the  European  Re\-search  Council  under  Grant  No.~681311  (QUESS), by the Academy of Finland through its Centre of Excellence in Quantum  Technology  (QTF)  (Grant  Nos.~312298  and~312300), by the Jane and Aatos Erkko Foundation, and
by the Technology Industries of Finland Centennial Foundation. It was also supported by the German Science Foundation (Grant Nos. AN336/11-1 and AN336/12-1), the Centre for Integrated Quantum Science and Technology (IQ$^{\rm ST}$), and the Zeiss Foundation under the Grant TQuant. The authors wish to acknowledge CSC~-- IT Center for Science, Finland, for computational resources.

\appendix

\section{Born-Markov master equations}



\label{sec:born-markov}

The Liouville--von Neumann equation which describes the dynamics of the density operator of the total system has the 
form
\begin{equation}
    \frac{d \hat \rho}{dt} = -\frac{i}{\hbar} \left[ \hat H, \hat \rho\right] .
\end{equation}
We eliminate the bath and system Hamiltonian from the above equation by moving to the interaction picture:
\begin{equation}
    \frac{d\hat{\tilde \rho}}{dt} = -\frac{i}{\hbar} \left[
        \hat{q}(t) \hat \xi(t), \hat{\tilde \rho}
        \right] ,
    \label{eq:interaction}
\end{equation}
where
\begin{equation}
    \hat \rho = \exp\left[-\frac{it}{\hbar}\left(\hat H_{\rm B} + \hat
    H_{\rm S}\right)\right]\hat{\tilde \rho}\exp\left[
        \frac{it}{\hbar} \left(\hat H_{\rm B} + \hat H_{\rm S}\right)
        \right],
\end{equation}
\begin{equation}
    \xi(t) = \sum\limits_k g_k \left(\hat b_k^\dag e^{i
    \omega_k t} + \hat b_k e^{-i\omega_k t}\right),
\end{equation}
and
\begin{equation}
    \hat q(t) = \sum_{nm} q_{nm} e^{i (\varepsilon_n - \varepsilon_m) t/{\hbar}
}|n\rangle \langle m| .
\end{equation}
Here $|n\rangle$ is the $n$-th eigenstate of the system Hamiltonian $\hat H_{\rm S}$, which corresponds to the eigenvalue $\varepsilon_n$. We follow the standard procedure and solve the density operator time-evolution iteratively from~(\ref{eq:interaction}). Taking the iteration to second order, we reach the formal expression
\begin{multline}
    \hat{\tilde \rho}(t) = \hat{\tilde \rho}(0) - \frac{i}{\hbar} \int\limits_0^t \left
    [\hat{q}(t') \hat \xi(t'), \hat{ \tilde \rho}(t')\right]\;dt' = \\
    \hat{\tilde \rho}(0) - \frac{i}{\hbar} \int\limits_0^t \left[\hat {q}(t') \hat \xi(t'),
    \hat{\tilde \rho}(0)\right]\;dt' -\\  \frac{1}{\hbar^2} \int\limits_0^t
    \int\limits_0^{t'} \left[\hat{q}(t') \hat \xi(t'), \left[
        \hat{q}(t'') \hat \xi(t''), \hat{\tilde \rho}(t'')\right]\right] \;dt'
    \;dt''
    \label{eq:dyson}
\end{multline}
with the intent of substituting a (possibly approximate) analytic solution for the inner integration.
For this purpose, one typically also assumes a factorized initial state
\begin{equation}
    \hat{\tilde \rho}(0) = \hat{\tilde \rho}_{\rm S}(0) \otimes \hat \rho_{\rm B},
\end{equation}
where $\hat{\tilde \rho}_{\rm S} = \mathop{\mathrm{Tr}_{\rm B}}\hat{\tilde \rho}$ and $\hat \rho_{\rm B} = \mathop{\mathrm{Tr}_{\rm S}}\hat{\tilde \rho}$ are the reduced density operators of the system and the bath, respectively. Moreover, if the coupling is weak it is reasonable to assume that the correlations of the bath decay on a timescale $\tau_{\rm B}$ much shorter than the relevant timescales in the interaction picture (relaxation and dephasing times). Assuming the differences $t-t'$ and $t-t''$ do not exceed this range, and assuming the rates are properly described as second-order effects, one can make the approximation
\begin{equation}
    \hat{\tilde \rho}(t-\tau) \approx \hat{\tilde \rho}_{\rm S}(t-\tau) \otimes \hat \rho_{\rm B}
    \label{eq:born-approximation}
\end{equation}
on the terms $\hat{\tilde\rho}$ appearing on the r.h.s.\ of  Eq. (\ref{eq:dyson}): Neglecting system-reservoir correlation effects at this point means neglecting effects of higher order in the interaction than second order. This constitutes the Born approximation on the coupled system-reservoir dynamics. Note that for factorizing initial states the Born-approximated dynamics cannot reveal even weak system-reservoir correlations unless the r.h.s.\ of Eq. (\ref{eq:dyson}) is evaluated in the full Liouville space.

\begin{figure}[tb]
\includegraphics[width = \linewidth]{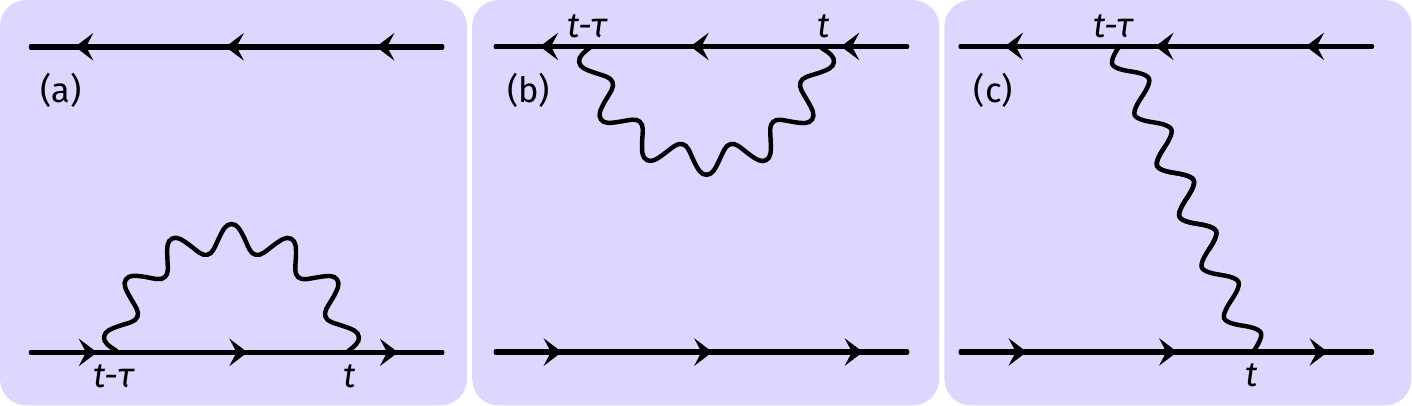}
\caption{\label{fig:born3}Schematic Feynman diagrams of processes implied in Markovian master equations. Solid lines represent system propagation, wiggly lines reservoir (de-)excitations. Diagrams (a) and (b) are self-energy-like corrections to the left/right application of the system Hamiltonian; (c) represents real emission or absorption. The partial trace implied by the reduced density matrix prevents the appearance of open in- or outgoing reservoir lines.}
\end{figure}

Substituting~(\ref{eq:born-approximation}) into~(\ref{eq:dyson}) and taking into account that $\langle \hat \xi \rangle = \mathop{\mathrm{Tr}_{\rm B}} \left(\hat \rho_{\rm B} \hat \xi\right) = 0$ we obtain
\begin{widetext}
\begin{multline}
    \frac{d \hat{\tilde \rho}_{\rm S}}{dt} = - \frac{1}{\hbar^2}
    \int\limits_0^t \left\{
        \left[\hat q(t) \hat q(t') \hat{\tilde \rho}_{\rm S}(t') - \hat q(t') \hat{\tilde
        \rho}(t') \hat q(t)\right]\left \langle \hat \xi(t) \hat \xi(t')\right
        \rangle +
        \left[
            \hat{\tilde \rho}_{\rm S}(t') \hat q(t') \hat q(t) - \hat q(t) \tilde{\hat
            \rho}_{\rm S}(t') \hat q(t') \right] \left\langle\hat \xi(t') \hat \xi(t)\right
        \rangle
        \right\}\;dt' = \\
    -\frac{1}{\hbar^2} \int\limits_0^t \left\{
        \left[\hat q(t) \hat q(t-\tau) \hat{\tilde \rho}_{\rm S}(t-\tau) - \hat
        q(t-\tau) \hat{\tilde
        \rho}(t-\tau) \hat q(t)\right]\left \langle \hat \xi(t) \hat \xi(t-\tau)\right
        \rangle + \right.  \\ \left.
        \left[
            \hat{\tilde \rho}_{\rm S}(t-\tau) \hat q(t-\tau) \hat q(t) - \hat q(t) \tilde{\hat
            \rho}_{\rm S}(t-\tau) \hat q(t-\tau) \right] \left\langle\hat \xi(t-\tau) \hat \xi(t)\right
        \rangle
        \right\}\; d\tau,
    \label{eq:born}
\end{multline}
\end{widetext}
where $\left\langle \hat \xi(t) \hat \xi(t - \tau) \right \rangle = \mathop{\mathrm{Tr}_{\rm B}}\left[\hat \xi(t) \hat \xi(t - \tau) \hat \rho_{\rm B} \right]$ is the bath correlation function. If the bath correlation function decays in a time scale $\tau_{\rm B}$, which is much shorter than any system time scale $\tau_{\rm S}$ , one can approximate it with function peaked at $\tau = 0$. In this limit, one typically makes the Markov approximation and assumes that $\hat{\tilde \rho}_{\rm S}(t - \tau) \sim \hat{\tilde \rho}_{\rm S}(t)$ in the region $\tau \lessapprox \tau_{\rm B}$ where the correlation function is appreciably different from zero. The time scale of the system in the interaction picture is again given by relaxation and dephasing, i.e. $\tau_{\rm S} \approx \kappa^{-1}$, where the rate $\kappa$ characterizes the strength of the bath coupling. However, one cannot make a similar approximation for the operator $\hat q(t - \tau)$ as it obtains an oscillating phase of the form $e^{i(\varepsilon_{n} - \varepsilon_m) t / \hbar}$. For a bath in a thermal equilibrium with a smooth, broad-band spectrum, the width of the correlation function is determined by the inverse temperature $\tau_{\rm B} \sim \hbar \beta$. Based on the above, the Markov approximation holds if $\hbar \kappa \beta \ll 1$. This is essentially the same condition used to justify the Born approximation. If the times $t$ under consideration obey $t \gg \hbar \beta$ , one can extend the limits of the integration in the Eq.~(\ref{eq:born}) to infinity, neglecting an initial slip which is typically insignificant.
Thus, one obtains
\begin{widetext}
\begin{multline}
    \frac{d \hat{\tilde \rho}_{\rm S}}{dt} = - \frac{1}{\hbar^2}
    \int\limits_0^{+\infty} \left\{
        \left[\hat q(t) \hat q(t-\tau) \hat{\tilde \rho}_{\rm S}(t) - \hat
        q(t-\tau) \hat{\tilde
        \rho}_{\rm S}(t) \hat q(t)\right]\left \langle \hat \xi(t) \hat \xi(t-\tau)\right
        \rangle + \right.  \\ \left.
        \left[
            \hat{\tilde \rho}_{\rm S}(t) \hat q(t-\tau) \hat q(t) - \hat q(t) \tilde{\hat
            \rho}_{\rm S}(t) \hat q(t-\tau) \right] \left\langle\hat \xi(t-\tau) \hat \xi(t)\right
        \rangle
        \right\}\; d\tau .
        \label{eq:born-markov}
\end{multline}
\end{widetext}
In the literature, this is referred to as the Born-Markov master equation of the reduced system density operator. In the Schr\"odinger picture, each of the individual terms describes a particular form of simultaneous propagation of system and reservoir between two interactions, graphically represented in the Feynman diagrams of Fig.\ \ref{fig:born3}. We remark that the justification of the Born approximation from the inequality $\hbar \kappa \beta \ll 1$ alone is not fully rigorous. For the typical case of reservoirs with a smooth and monotonically rising density of states, higher-order corrections seem to be irrelevant at temperatures low enough compared to energy splittings of the system~\cite{napol94}.

Performing integration over $\tau$ in the equation~(\ref{eq:born-markov}) and neglecting the correction to the coherent part of the Hamiltonian (the Lamb shift) one obtains the Redfield master equation. By making a transformation back to the Schr\"odinger picture, the Redfield equation can be written in the eigenbasis of the Hamiltonian $\hat H_{\rm S}$ as 
\begin{equation}
    \dot \rho_{jk} = i \omega_{jk} \rho_{jk} - \sum\limits_{lm} R_{jklm} \rho_{lm} ,
    \label{eq:redfield}
\end{equation}
where
\begin{multline}
    R_{jklm} = \frac{1}{2\hbar^2} \left\{
    \delta_{km} \sum\limits_n S(\omega_{nl}) q_{jn} q_{nl} + \right. \\  \delta_{jl} \sum\limits_n S(-\omega_{mn}) q_{mn} q_{nk} - \\ \left. \phantom{\sum\limits_n} \left[S(\omega_{jl}) + S(-\omega_{mk})\right]q_{jl} q_{mk}
    \right\} 
\end{multline}
and $q_{nm} = \langle n | \hat q | m \rangle$, $\omega_{nm} = \omega_m - \omega_n$, $\hat H_S | n\rangle = \hbar \omega_n |n \rangle$ and $S(\omega)$ is the Fourier image of the bath correlation function:
\begin{multline}
    S(\omega) = \int\limits_{-\infty}^{+\infty}\left\langle \hat \xi(t) \hat \xi(t - \tau) \right \rangle e^{i \omega \tau}\;d\tau =  \frac{2\hbar J(\omega)}{1 - e^{-\hbar \beta \omega}}.
\end{multline}
The last equation holds for the bath in the thermal equilibrium.

The Lindblad equation is obtained from the Redfield equation written in the interaction picture using the secular approximation. For a system with a  non-degenerate spectrum the Lindblad equation, restricted to diagonal states in the eigenbasis of the Hamiltonian $\hat H_{\rm S}$, reads
\begin{equation}
    \dot \rho_{nn} = \sum\limits_{m} \left[\Gamma_{m\to n} \rho_{mm} - \Gamma_{n \to m} \rho_{nn}\right] ,
\end{equation}
where we have denoted reservoir-induced transition rates between the eigenstates by
\begin{equation}
    \Gamma_{m\to n} = \frac{|q_{nm}|^2}{\hbar^2} S(-\omega_{mn}) \ .
\end{equation}
Thus, the diagonal density matrix elements are decoupled from the off-diagonal ones. The temporal evolution of the off-diagonal terms can be also calculated and we find that they approach the steady state as
\begin{equation}
    \dot \rho_{nm} = i\omega_{nm} \rho_{nm} -\left(\gamma_{nm} + \gamma_{nm}^\phi\right) \rho_{nm} \ ,
\end{equation}
where the losses in the phase coherence are caused by relaxation:
\begin{equation}
    \gamma_{nm} = \frac{1}{2} \sum\limits_{k\ne n,m} \left[\Gamma_{n\to k} + \Gamma_{m \to k} \right]
\end{equation}
and pure dephasing:
\begin{equation}
    \gamma_{nm}^\phi = \frac{1}{2\hbar^2} S(0) [q_{nn} - q_{mm}]^2
\end{equation}
between the states $|n\rangle$ and $|m\rangle$.

Alternatively, one may represent the time dependence of the interaction picture operators $\hat{q}(t)$ and $\hat{q}(t-\tau)$ in Eq.~(\ref{eq:born-markov}) through eigenoperators~\cite{breuerpetruccione02} of the superoperator $[H_{\rm S},\cdot]$. For a harmonic oscillator for example, these are the raising and lowering operators. Thus, the secular approximation leads to neglecting the terms in Eq.~(\ref{eq:born-markov}) 
that oscillate at the system frequencies, and keeping those terms which are constants with respect to $t$. This procedure is meaningful only if the damping is weak enough to permit a significant number of oscillation cycles between emission events, i.\,e., if the relaxation and dephasing rates are lower than all transition frequencies. This approach leads directly to the commonly used general form of a Lindblad master equation,
\begin{equation}
\frac{d\hat{\tilde\rho}_{\rm S}}{dt}=-{i\over\hbar}[H,\hat{\tilde\rho}_{\rm S}]+\sum_\alpha \gamma_\alpha\left(L_\alpha\hat{\tilde\rho}_{\rm S} L_\alpha^\dagger -\frac{1}{2} \left\{L_\alpha^\dagger L_\alpha, \hat{\tilde\rho}_{\rm S}\right\} \right).
\end{equation}
In the case of weakly interacting qubits, the introduction of yet another small parameter $g$ complicates both the determination of eigenoperators and the application of the secular approximation. If $g$ is small, the assumptions used in the construction of the Lindblad master equation are easily violated. However, in case $g$ is even smaller than relaxation and dephasing rates, one may altogether neglect it in the construction of the Lindbladian, which leads to the \emph{local} Lindblad approach.

\section{Stochastic Liouville equations}

\label{sec:stochastic-liouville}
The time evolution is solved from the stochastic Liouville-von Neumann (SLN) equation which can be written into the form
\begin{equation}
    i\hbar \frac{d\hat \rho_{\rm S}}{dt} = [\hat H_{\rm S}, \hat \rho_{\rm S}] - \zeta [\hat q, \hat \rho_{\rm S}] - \frac{\hbar}{2} \nu \{\hat q, \hat \rho_{\rm S}\} \ ,
\end{equation}
where $\zeta$ and $\nu$ are complex noise terms that arise from exact treatment of the coupling in the path-integral formalism. Together with the anti-commutator, these terms result in non-unitary time evolution for individual samples, i.e. realizations of the noise terms. However, by making a stochastic average, the non-hermitian parts parts of the density operator vanish. Also the trace of the density operator is unity on average. The noise terms obey the correlation functions
\begin{equation}
    \langle \zeta(t) \zeta(t') \rangle = \mathop{\mathrm{Re}} L(t - t') \ ,
\end{equation}
\begin{multline}
    \langle \zeta(t) \nu(t') \rangle = \\ \frac{2i}{\hbar} \Theta(t - t') \mathop{\mathrm{Im}} L(t - t') + i \mu \delta(t - t') = \\ -i\chi_{\rm R}(t - t') + i \mu \delta(t - t') \ ,
\end{multline}
\begin{equation}
    \langle \nu(t) \nu(t') \rangle = 0 \ .
\end{equation}
Above, the bath correlation function
\begin{multline}
    L(t - t') =\\ \frac{\hbar}{\pi} \int\limits_0^{+\infty}d\omega\;J(\omega) \left\{
    \coth\left(\frac{\hbar \beta \omega}{2}\right) \cos[\omega(t - t')] -\right . \\ \left.\phantom{\frac{\beta}{2}} i \sin[\omega(t - t')]
    \right\},
\end{multline}
and we have defined the classical response function
\begin{equation}
    \chi_{\rm R}(t) = -\frac{2}{\hbar}\Theta(t) \mathop{\mathrm{Im}} L(t) \ ,
\end{equation}
and
\begin{equation}
    \mu = \int_{-\infty}^{+\infty}dt\;\chi_{\rm R}(t) \ .
\end{equation}
In the case of the ohmic spectral density with a Drude cutoff, defined in~(\ref{eq:ohmic})
where the cutoff frequency $\omega_c$ is much larger than any other frequency in the system, one can write SLN equation into the form of the stochastic Liouville equation with dissipation (SLED):
\begin{multline}
    i \hbar \frac{d \hat \rho}{dt} = \left[ \hat H_{\rm S}, \hat \rho \right] - \frac{i\eta}{\hbar \beta} \left[
    \hat q, \left[ \hat q, \hat \rho \right]
    \right] + \\  \frac{i\eta}{2 \hbar} \left[
        \hat q, \left\{
            \left[ \hat H_{\rm S}, \hat q\right], \hat \rho
        \right\}
    \right] - \zeta \left[\hat q, \hat \rho_{\rm S}\right] \ ,
\end{multline}
where $\zeta(t)$ is a stochastic Gaussian process with the following correlation function:
\begin{multline}
    \langle \zeta(t) \zeta(t')\rangle =\\ \frac{\hbar}{\pi} \int\limits_0^{+\infty}d\omega\; J(\omega) \left[
        \coth\left(\frac{\hbar \beta \omega}{2}\right) - \frac{2}{\hbar \beta \omega}
    \right] \cos[\omega (t - t')] \ .
\end{multline}
Notice that the assumption of the large cutoff frequency is used only in the noise process $\nu$; the cutoff frequency is still present in the autocorrelation function of $\zeta$. As a consequence of this approximation, the two complex noise terms in the SLN equation have been reduced into a deterministic part and a single real-valued noise term.

\bibliography{bibliography}

\end{document}


\preprint{APS/123-QED}

\title{Supplementary Material:\\
Validity of Born--Markov master equations for single and two-qubit systems} 

\author{Vasilii Vadimov}
\affiliation{QCD Labs and MSP group, QTF Centre of Excellence, Department of Applied Physics, Aalto University, P.O. Box 15100, FI-00076 Aalto, Finland}
\affiliation{Institute for Physics of Microstructures, Russian Academy of Sciences, 603950 Nizhny Novgorod, GSP-105, Russia}
\author{Jani Tuorila}
\affiliation{QCD Labs and MSP group, QTF Centre of Excellence, Department of Applied Physics, Aalto University, P.O. Box 15100, FI-00076 Aalto, Finland}
\affiliation{IQM, Keilaranta 19, FI-02150 Espoo, Finland}
\author{Tuure Orell}
\affiliation{Nano and Molecular Materials Research Unit, University of Oulu, P.O. Box 3000, FI-90014, Finland}
\author{J\"urgen Stockburger}
\affiliation{Institute for Complex Quantum Systems and IQST, University of Ulm, 89069 Ulm, Germany}
\author{Tapio Ala-Nissila}
\affiliation{QCD Labs and MSP group, QTF Centre of Excellence, Department of Applied Physics, Aalto University, P.O. Box 15100, FI-00076 Aalto, Finland}
\affiliation{Interdisciplinary Centre for Mathematical Modelling, Department of Mathematical Sciences, Loughborough University, Loughborough, Leicestershire LE11 3TU, UK}
\author{Joachim Ankerhold}
\affiliation{Institute for Complex Quantum Systems and IQST, University of Ulm, 89069 Ulm, Germany}
\author{Mikko M\"ott\"onen}
\affiliation{QCD Labs and MSP group, QTF Centre of Excellence, Department of Applied Physics, Aalto University, P.O. Box 15100, FI-00076 Aalto, Finland}
\affiliation{VTT Technical Research Centre of Finland Ltd., QTF Center of Excellence, P.O. Box 1000, FI-02044 VTT, Finland}

\date{\today}
\begin{abstract}{
This Supplementary contains details of parameter optimization for Born--Markov master equations describing a single qubit coupled to a heat bath and two mutually coupled qubits, one of which is coupled to a bath. The reliability of the fitting procedure and the corresponding fitting parameters are also discussed.
}
\end{abstract}
          
\maketitle

\section{Optimization parameters for a single-qubit system}

\label{sec:optimization-params-single}

In this section we show fitting parameters $\kappa^\mathrm{opt}$, $\beta^\mathrm{opt}$, and $\omega_q^\mathrm{opt}$ for the optimized BM master equations for a single-qubit case. The parameters remain reasonable for weak coupling below $\kappa < 0.3\omega_q$. In the strong coupling limit the optimization procedure appears to be somewhat unreliable and yields unphysical values for some of the parameters. However, even in the weak coupling regime $\kappa \sim 0.1 \omega_q$, at low temperatures $\hbar \beta \omega_q \gg 1$ the fitting parameter $\beta^\mathrm{opt}$ significantly deviates from the original $\beta$. This means that even if it is possible to fit the dynamics of the system with BM master equations, the values of the fitting parameters may significantly deviate from the real ones.

\begin{figure}[!ht]
\includegraphics[width=\linewidth]{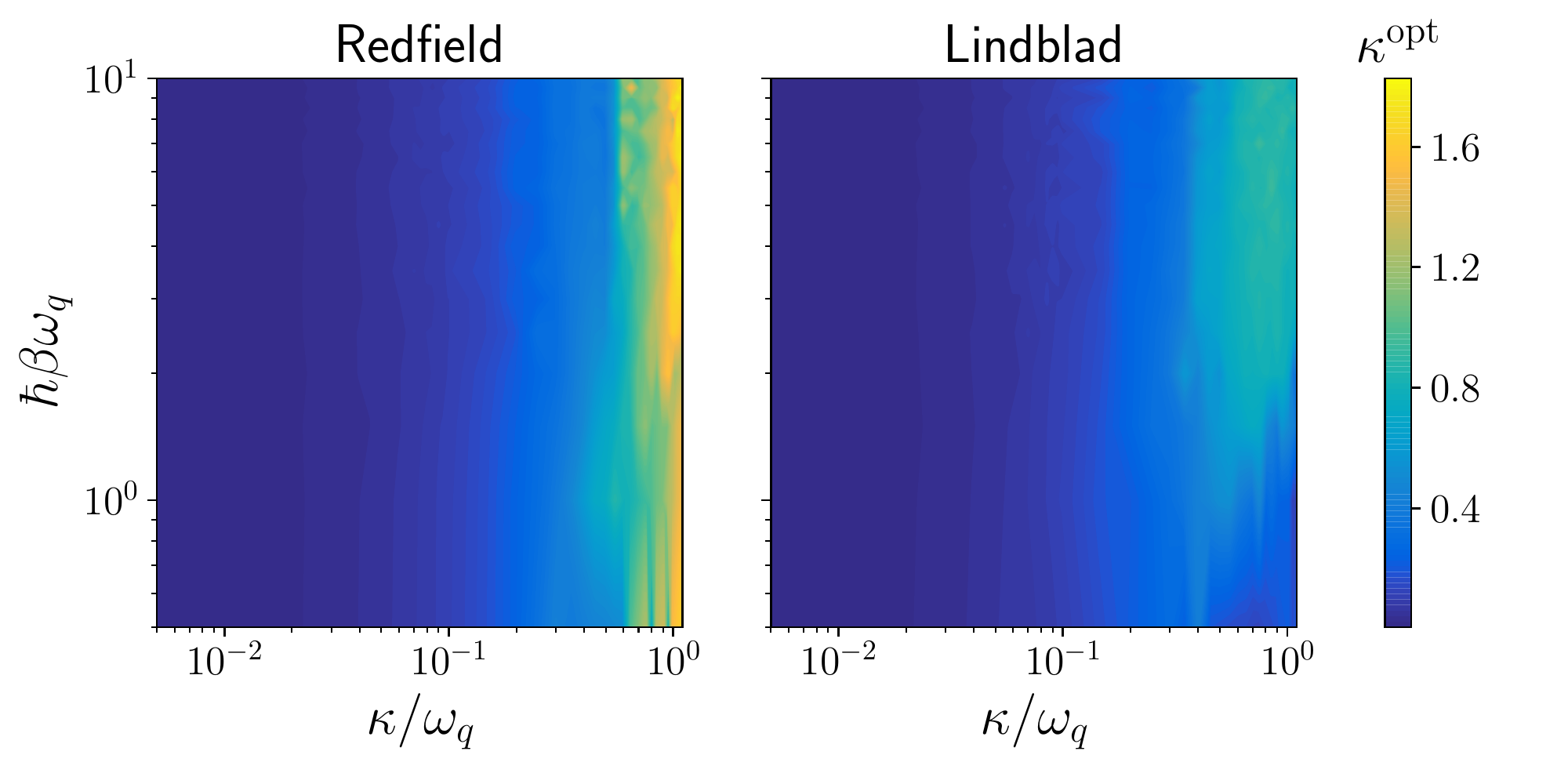}
\caption{Optimized values of $\kappa^\mathrm{opt}$ as a function of non-optimized values of $\kappa$ and $\beta$. The optimization is done by minimizing the value of $\Delta_\mathrm{max}$ as a function of $\kappa$, $\beta$, and $\omega_{\rm q}$.\label{fig:optkappa}}
\end{figure}

\begin{figure}[!ht]
\includegraphics[width=\linewidth]{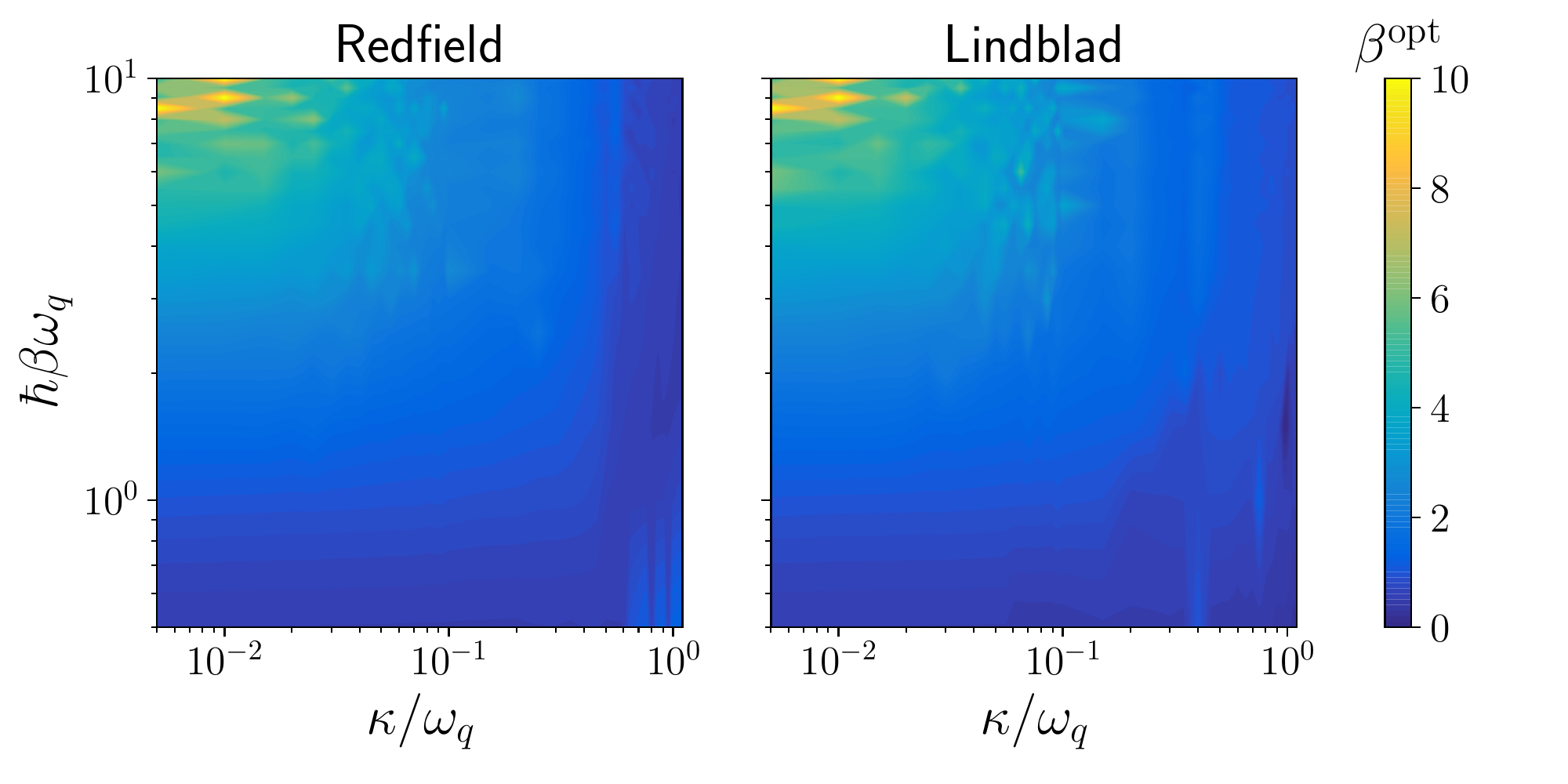}
\caption{Optimized values of $\beta^\mathrm{opt}$ as a function of non-optimized values of $\kappa$ and $\beta$. The optimization is done by minimizing the value of $\Delta_\mathrm{max}$ as a function of $\kappa$, $\beta$, and $\omega_{\rm q}$.}\label{fig:optbeta}
\end{figure}

\begin{figure}[!ht]
\includegraphics[width=\linewidth]{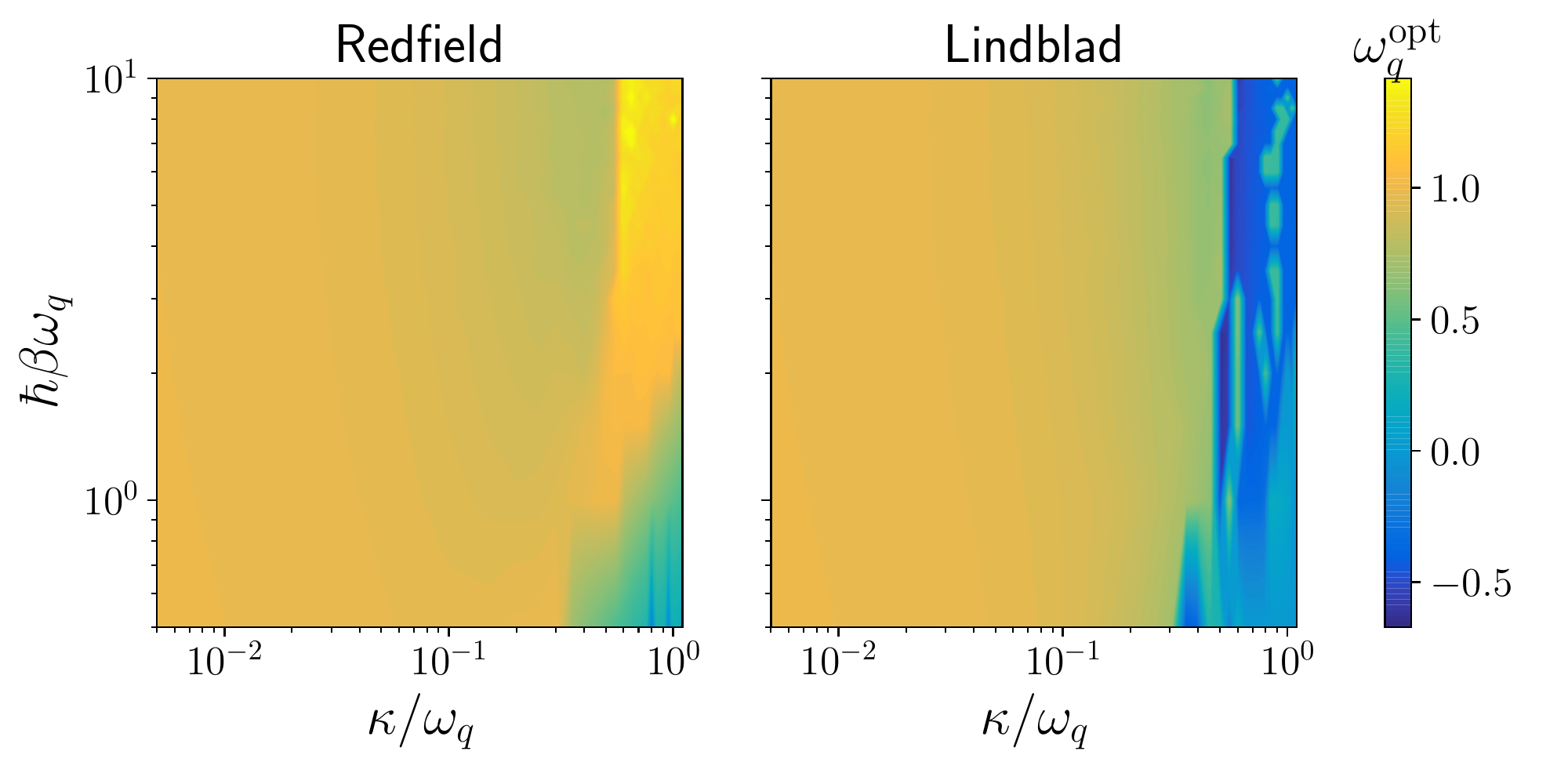}
\caption{Optimized values of $\omega_q^\mathrm{opt}$ as a function of non-optimized values of $\kappa$ and $\beta$. The optimization is done by minimizing the value of $\Delta_\mathrm{max}$ as a function of $\kappa$, $\beta$, and $\omega_{\rm q}$.}\label{fig:optomega}
\end{figure}

\section{Optimization parameters for a two-qubit system}
\label{sec:optimization-params-two}

Here we show the optimization parameters $\kappa^\mathrm{opt}$ and $\beta^\mathrm{opt}$, but instead of explicit plotting the parameters $a_1$, $a_2$ and $a_3$ which have no direct physical meaning we show the corrections to the transition frequencies between the neighboring levels of the system. More specifically, we plot the following quantities:
\begin{equation}
    \delta \omega_{jk}^\mathrm{opt} = \omega_{jk}^\mathrm{opt} - \omega_{jk},
\end{equation}
where $\omega_{jk}$ and $\omega_{jk}'$ correspond to the bare and Lamb-shifted transition frequencies
\begin{gather}
    \hbar \omega_{jk} = E_{k} - E{j}; \\
    \hbar \omega_{jk}^\mathrm{opt} = a_1 (E_k - E_j) + a_2 (E_k^2 - E_j^2) + a_3(E_k^3 - E_j^3),
\end{gather}
and $E_j$ are the energy levels of the bare Hamiltonian of the two-qubit system. It is enough to consider only adjacent transitions $k = j + 1$ since the other transition frequencies come as combinations of the ones considered.

In the case of high ($\hbar \beta \omega_1 = 0.1$) and intermediate ($\hbar \beta \omega_1 = 1$) temperatures the optimization procedure seems to be quite stable and gives reasonable values for the parameters. However, the fitted parameters sometimes significantly differ from the bare ones, especially for approaches based on Lindblad equations in the case of strong coupling and lower temperature. At low temperature ($\hbar \beta \omega_1 = 5$) stability of the optimization breaks down which is most likely caused by the presence of several local minima in the optimized function $\Delta_\mathrm{max}$. Even in cases where the fit parameters are reasonable they still usually deviate from the bare ones when one approaches the strong coupling regime. This means that the optimization gives correct parameters of the microscopic model only in the weak-coupling limit.

\begin{figure*}[!ht]
\includegraphics[width=\linewidth]{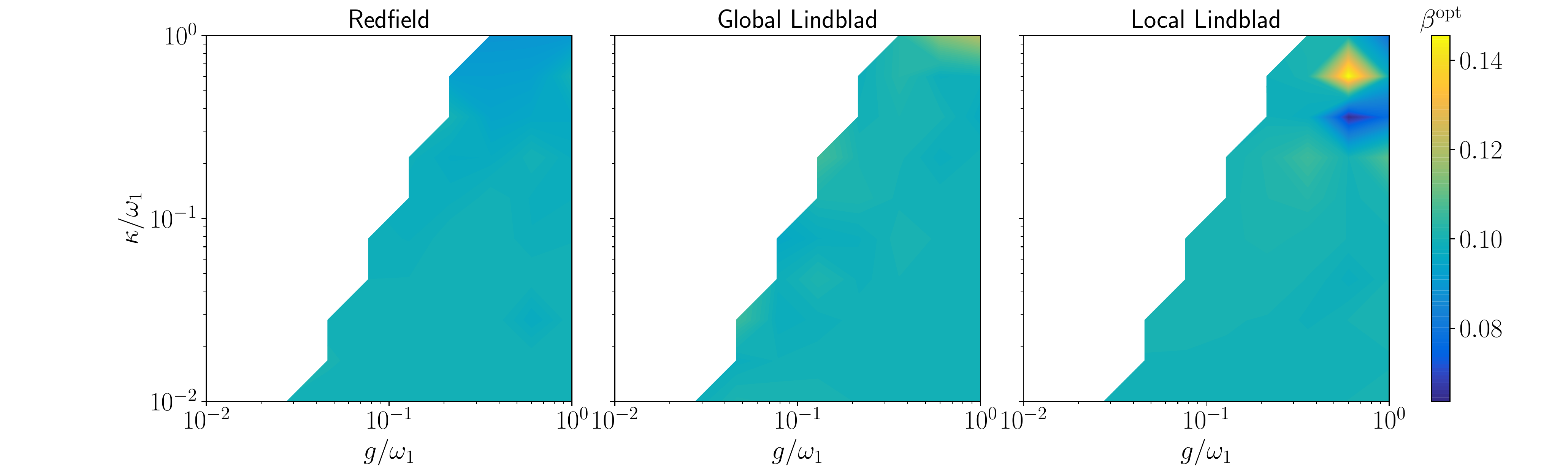}
\caption{Optimized values of $\beta^\mathrm{opt}$ as a function of non-optimized values of $g$ and $\kappa$ at high bare temperature $\hbar \beta \omega_1 = 0.1$. The optimization is done by minimizing the value of $\Delta_\mathrm{max}$ as a function of $\kappa$, $\beta$, $a_1$, $a_2$, and $a_3$.}\label{fig:opt-beta-high}
\end{figure*}

\begin{figure*}[!ht]
\includegraphics[width=\linewidth]{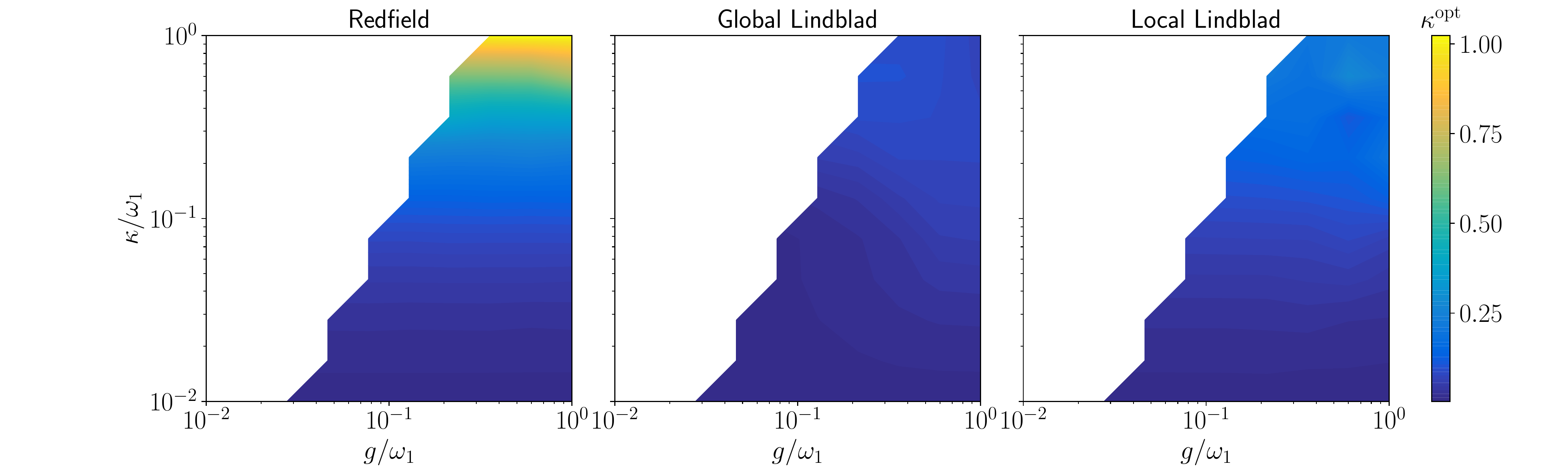}
\caption{Optimized values of $\kappa^\mathrm{opt}$ as a function of non-optimized values of $g$ and $\kappa$ at high bare temperature $\hbar \beta \omega_1 = 0.1$. The optimization is done by minimizing the value of $\Delta_\mathrm{max}$ as a function of $\kappa$, $\beta$, $g$,  $\omega_1$ and $\omega_2$.}
\end{figure*}

\begin{figure*}[!ht]
\includegraphics[width=\linewidth]{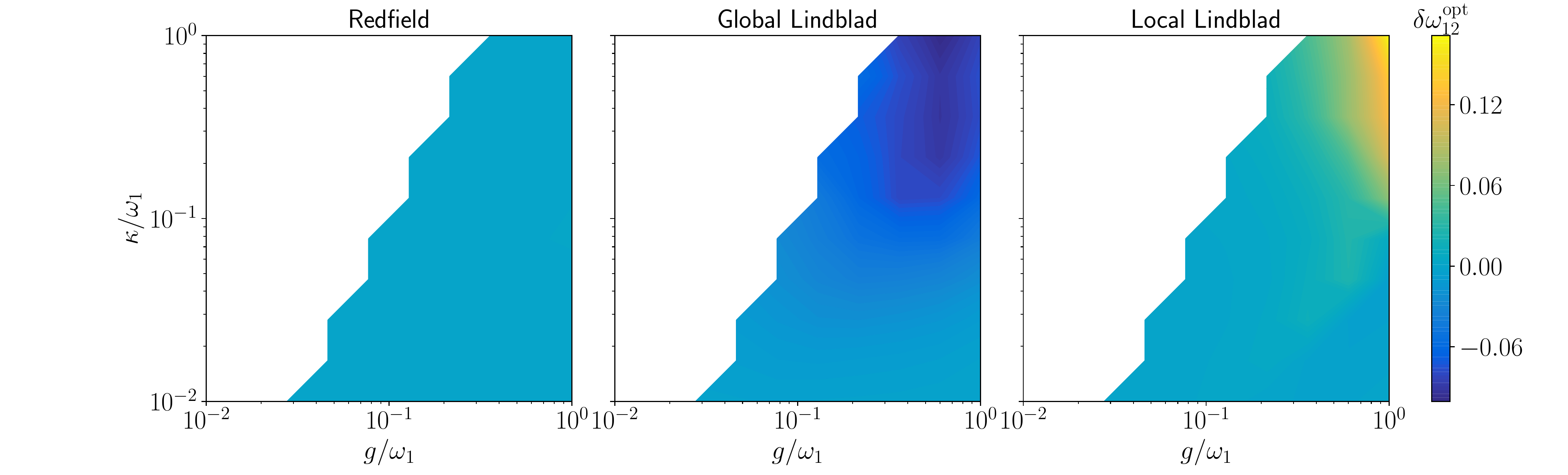}
\caption{Optimized values of $\delta \omega_{12}^\mathrm{opt}$ as a function of non-optimized values of $g$ and $\kappa$ at high bare temperature $\hbar \beta \omega_1 = 0.1$. The optimization is done by minimizing the value of $\Delta_\mathrm{max}$ as a function of $\kappa$, $\beta$, $a_1$, $a_2$, and $a_3$.}
\end{figure*}

\begin{figure*}[!ht]
\includegraphics[width=\linewidth]{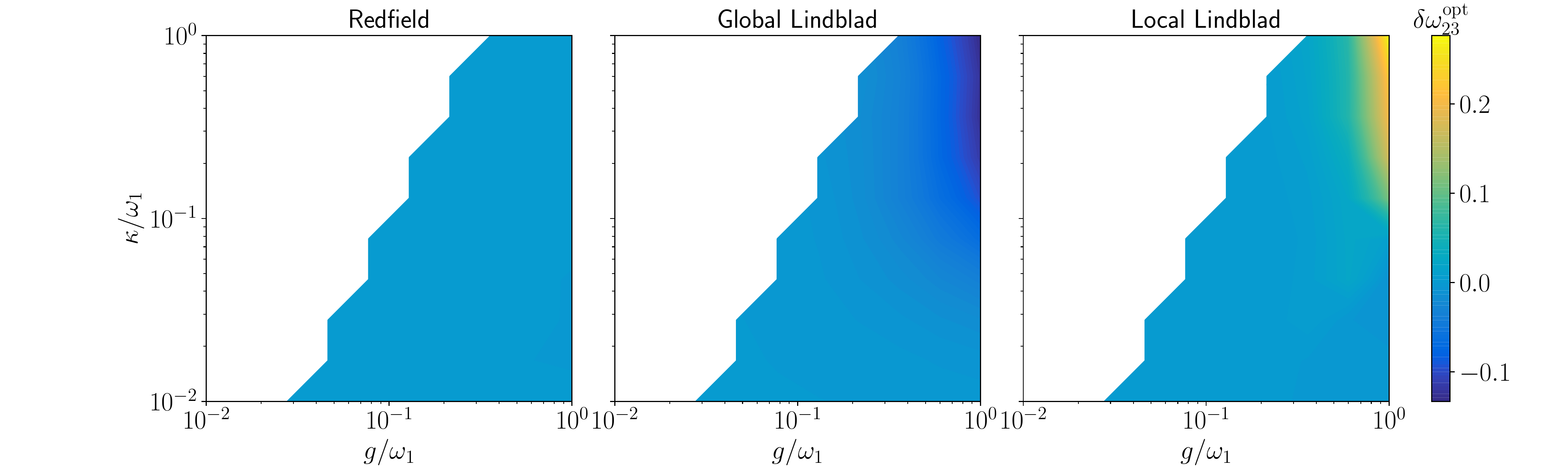}
\caption{Optimized values of $\delta \omega_{23}^\mathrm{opt}$ as a function of non-optimized values of $g$ and $\kappa$ at high bare temperature $\hbar \beta \omega_1 = 0.1$. The optimization is done by minimizing the value of $\Delta_\mathrm{max}$ as a function of $\kappa$, $\beta$, $a_1$, $a_2$, and $a_3$.}
\end{figure*}

\begin{figure*}[!ht]
\includegraphics[width=\linewidth]{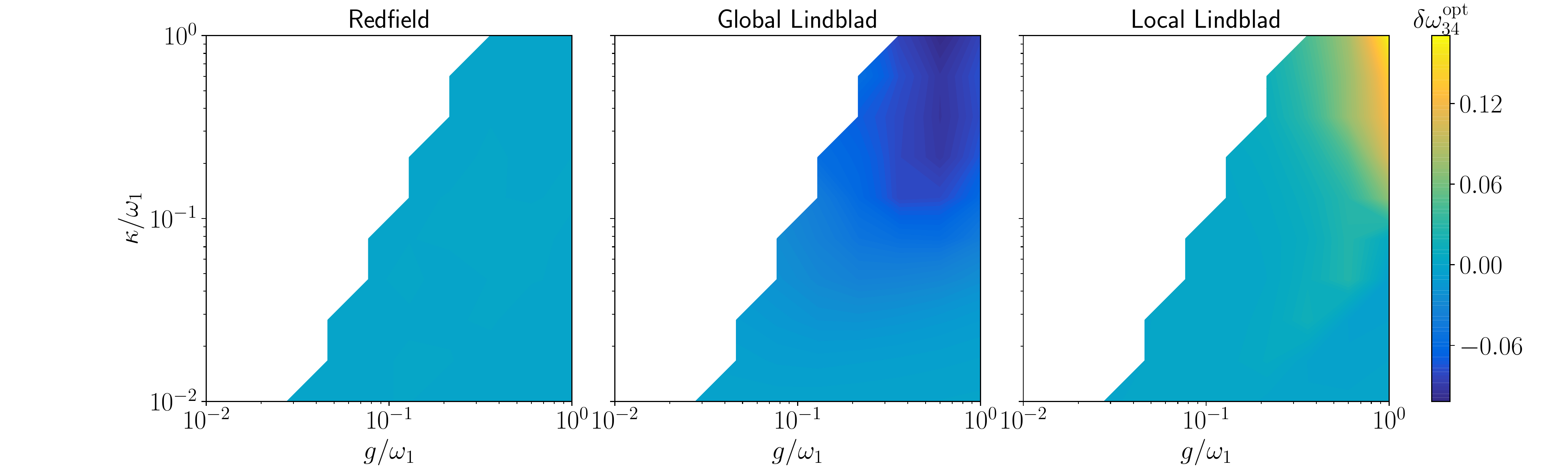}
\caption{Optimized values of $\delta \omega_{34}^\mathrm{opt}$ as a function of non-optimized values of $g$ and $\kappa$ at high bare temperature $\hbar \beta \omega_1 = 0.1$. The optimization is done by minimizing the value of $\Delta_\mathrm{max}$ as a function of $\kappa$, $\beta$, $a_1$, $a_2$, and $a_3$.}
\end{figure*}

\begin{figure*}[!ht]
\includegraphics[width=\linewidth]{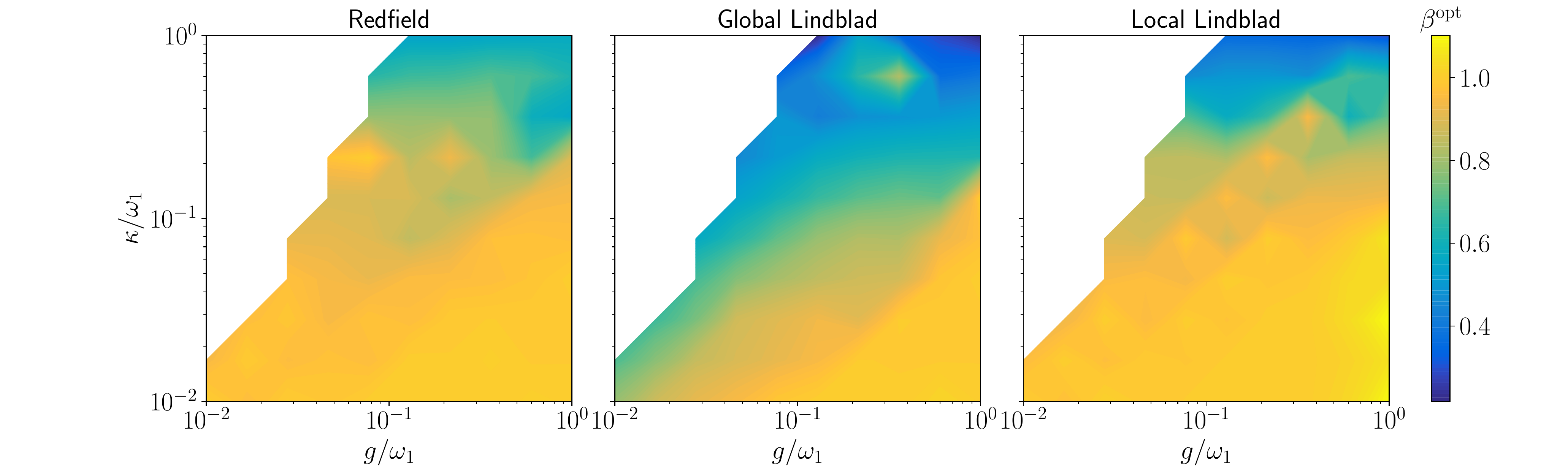}
\caption{Optimized values of $\beta^\mathrm{opt}$ as a function of non-optimized values of $g$ and $\kappa$ at intermediate bare temperature $\hbar \beta \omega_1 = 1$. The optimization is done by minimizing the  value of $\Delta_\mathrm{max}$ as a function of $\kappa$, $\beta$, $a_1$, $a_2$, and $a_3$.}
\end{figure*}

\begin{figure*}[!ht]
\includegraphics[width=\linewidth]{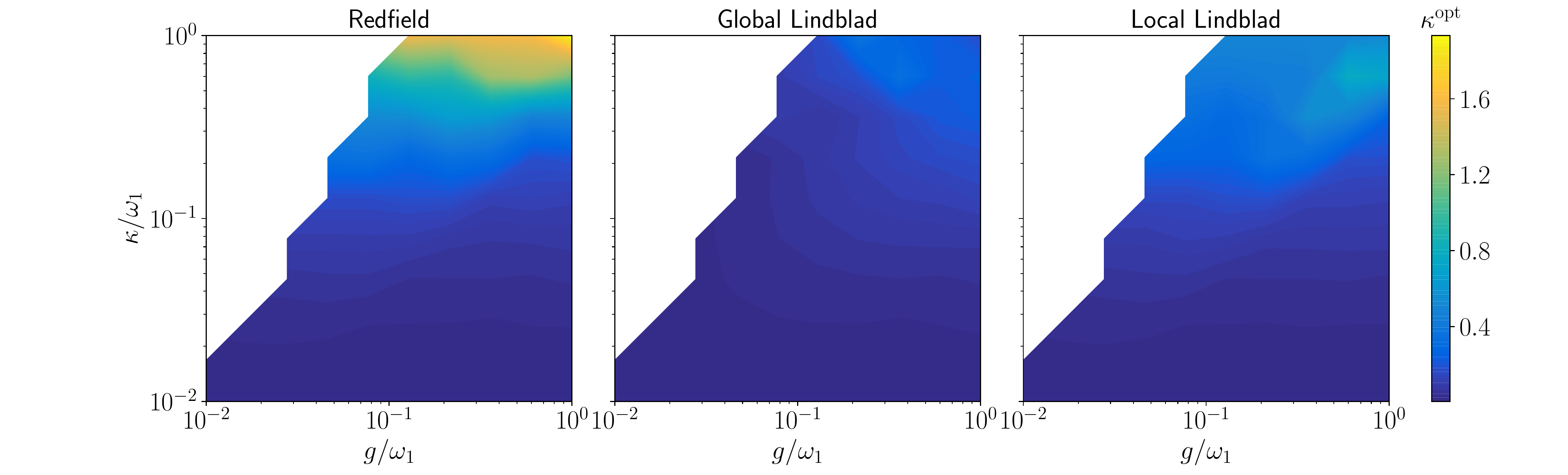}
\caption{Optimized values of $\kappa^\mathrm{opt}$ as a function of non-optimized values of $g$ and $\kappa$ at intermediate bare temperature $\hbar \beta \omega_1 = 1$. The optimization is done by minimizing the  value of $\Delta_\mathrm{max}$ as a function of $\kappa$, $\beta$, $a_1$, $a_2$, and $a_3$.}
\end{figure*}

\begin{figure*}[!ht]
\includegraphics[width=\linewidth]{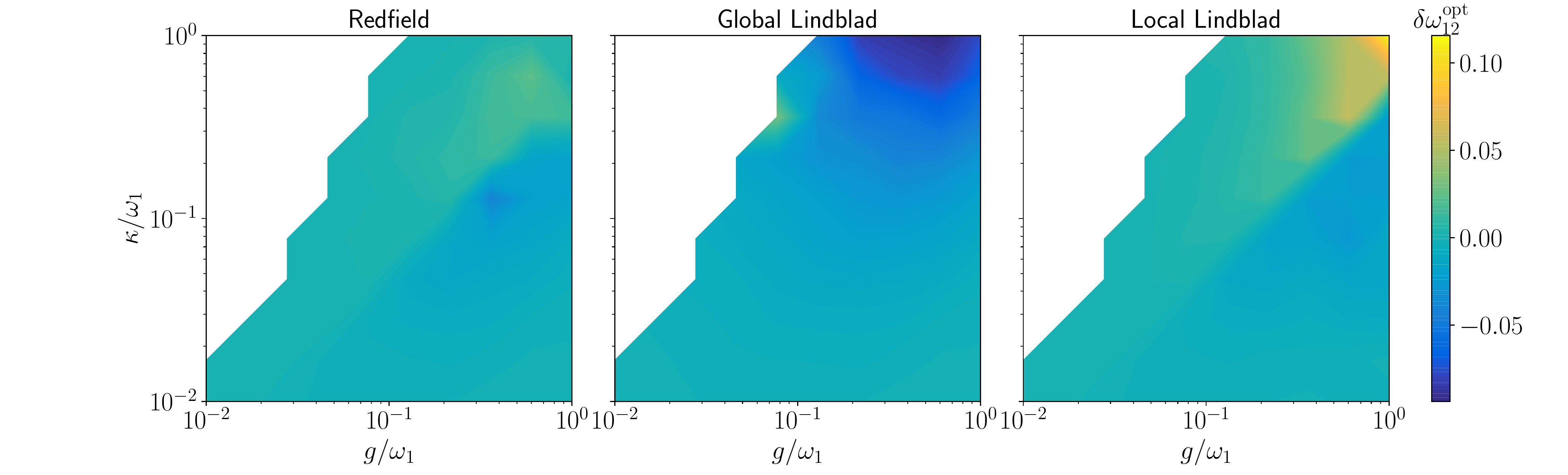}
\caption{Optimized values of $\delta \omega_{12}^\mathrm{opt}$ as a function of non-optimized values of $g$ and $\kappa$ at intermediate bare temperature $\hbar \beta \omega_1 = 1$. The optimization is done by minimizing the value of $\Delta_\mathrm{max}$ as a function of $\kappa$, $\beta$, $a_1$, $a_2$, and $a_3$.}
\end{figure*}

\begin{figure*}[!ht]
\includegraphics[width=\linewidth]{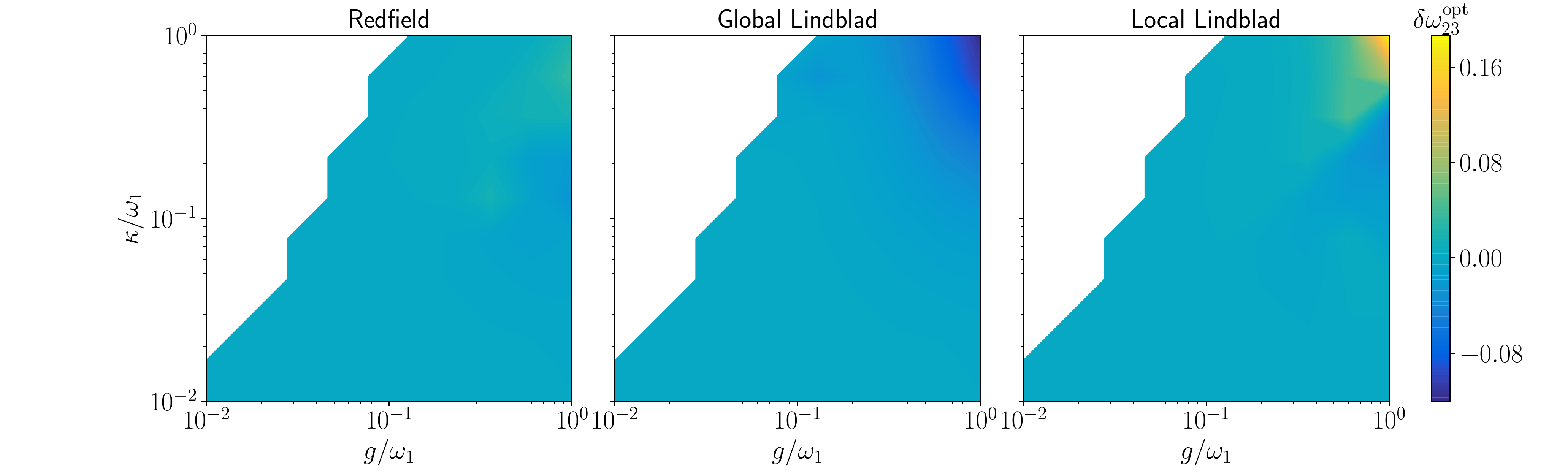}
\caption{Optimized values of $\delta \omega_{23}^\mathrm{opt}$ as a function of non-optimized values of $g$ and $\kappa$ at high intermediate temperature $\hbar \beta \omega_1 = 1$. The optimization is done by minimizing the  value of $\Delta_\mathrm{max}$ as a function of $\kappa$, $\beta$, $a_1$, $a_2$, and $a_3$.}
\end{figure*}

\begin{figure*}[!ht]
\includegraphics[width=\linewidth]{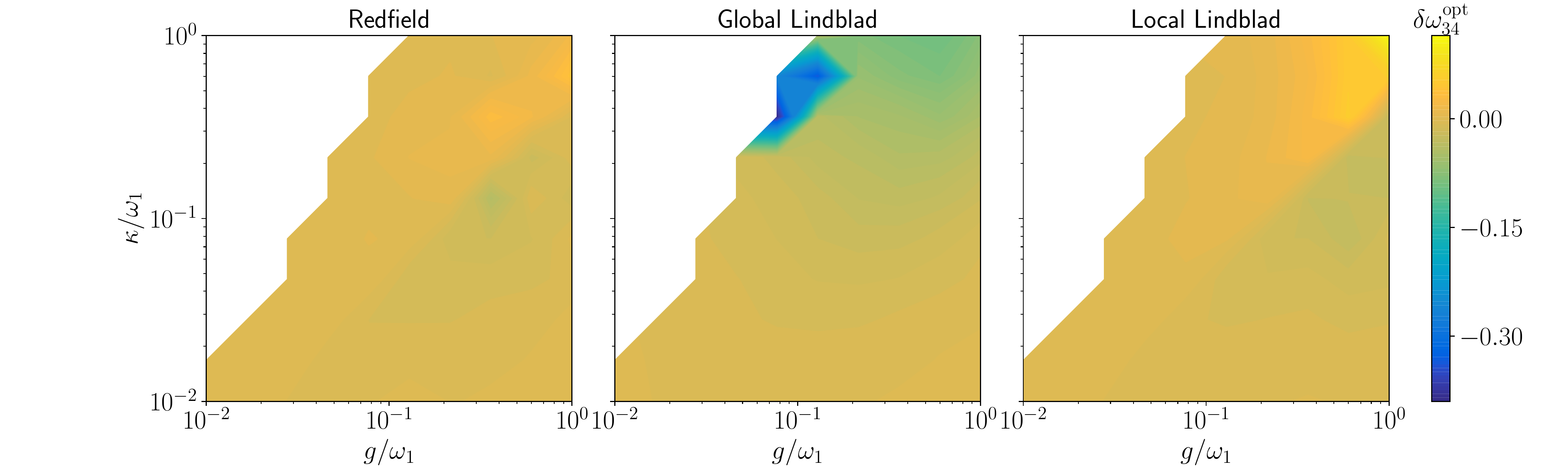}
\caption{Optimized values of $\delta \omega_{34}^\mathrm{opt}$ as a function of non-optimized values of $g$ and $\kappa$ at intermediate bare temperature $\hbar \beta \omega_1 = 1$. The optimization is done by minimizing the  value of $\Delta_\mathrm{max}$ as a function of $\kappa$, $\beta$, $a_1$, $a_2$, and $a_3$.}
\end{figure*}

\begin{figure*}[!ht]
\includegraphics[width=\linewidth]{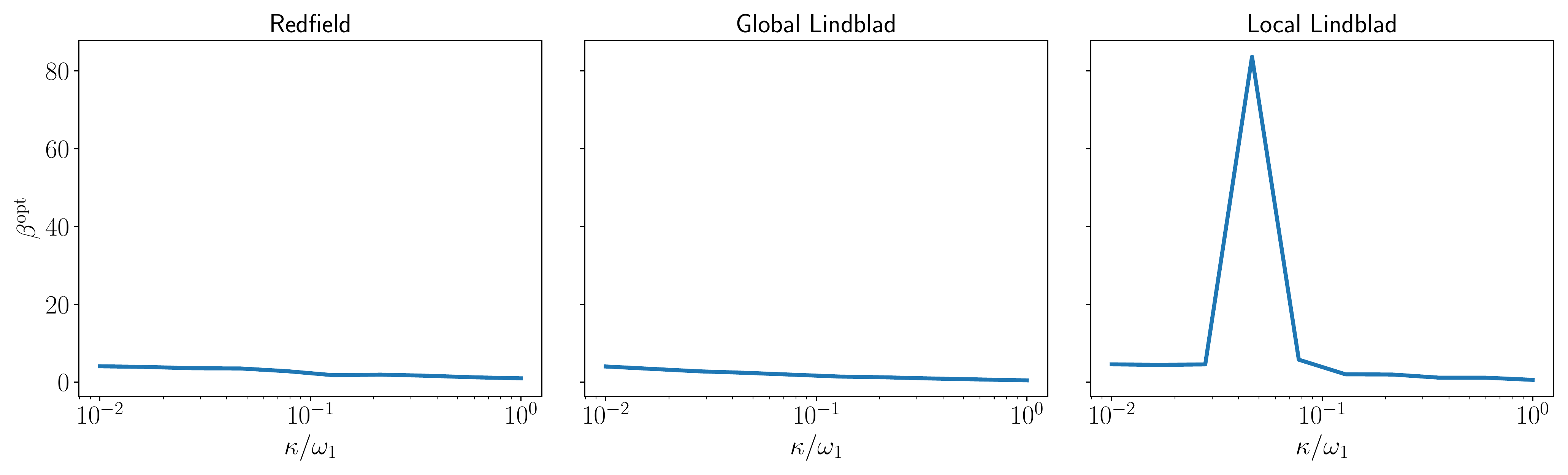}
\caption{Optimized values of $\beta^\mathrm{opt}$ as a function of non-optimized value of $\kappa$ at low bare temperature $\hbar \beta \omega_1 = 5$, $g = 0.1 \omega_1$. The optimization is done by minimizing the  value of $\Delta_\mathrm{max}$ as a function of $\kappa$, $\beta$, $a_1$, $a_2$, and $a_3$.}
\end{figure*}

\begin{figure*}[!ht]
\includegraphics[width=\linewidth]{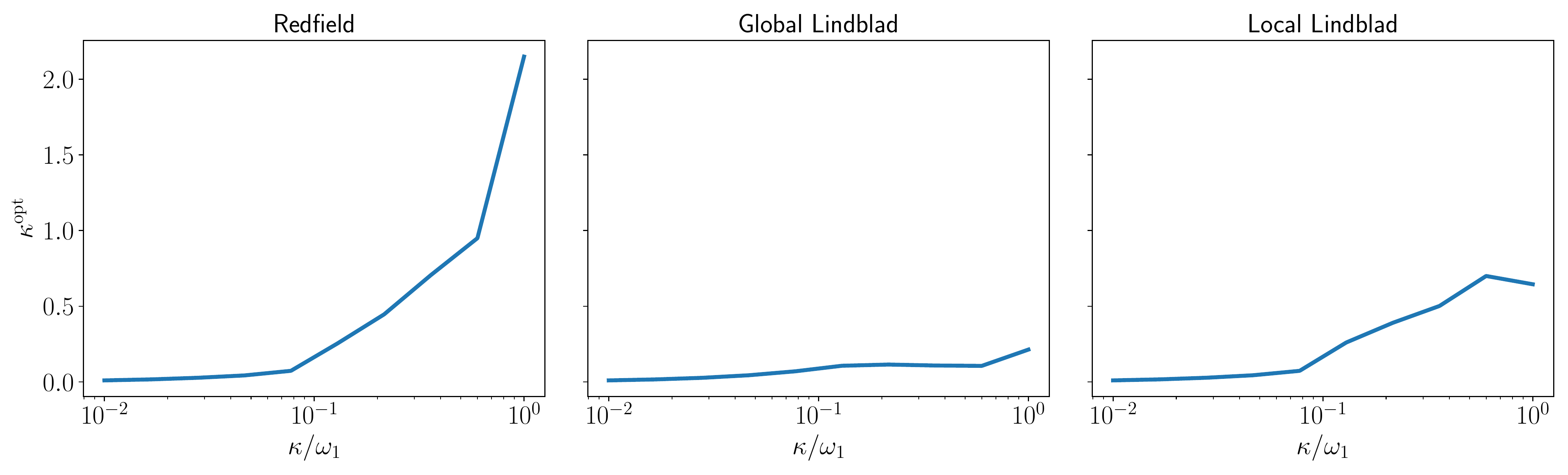}
\caption{Optimized values of $\kappa^\mathrm{opt}$  as a function of non-optimized value of $\kappa$ at low bare temperature $\hbar \beta \omega_1 = 5$, $g = 0.1 \omega_1$. The optimization is done by minimizing the  value of $\Delta_\mathrm{max}$ as a function of $\kappa$, $\beta$, $a_1$, $a_2$, and $a_3$.}
\end{figure*}

\begin{figure*}[!ht]
\includegraphics[width=\linewidth]{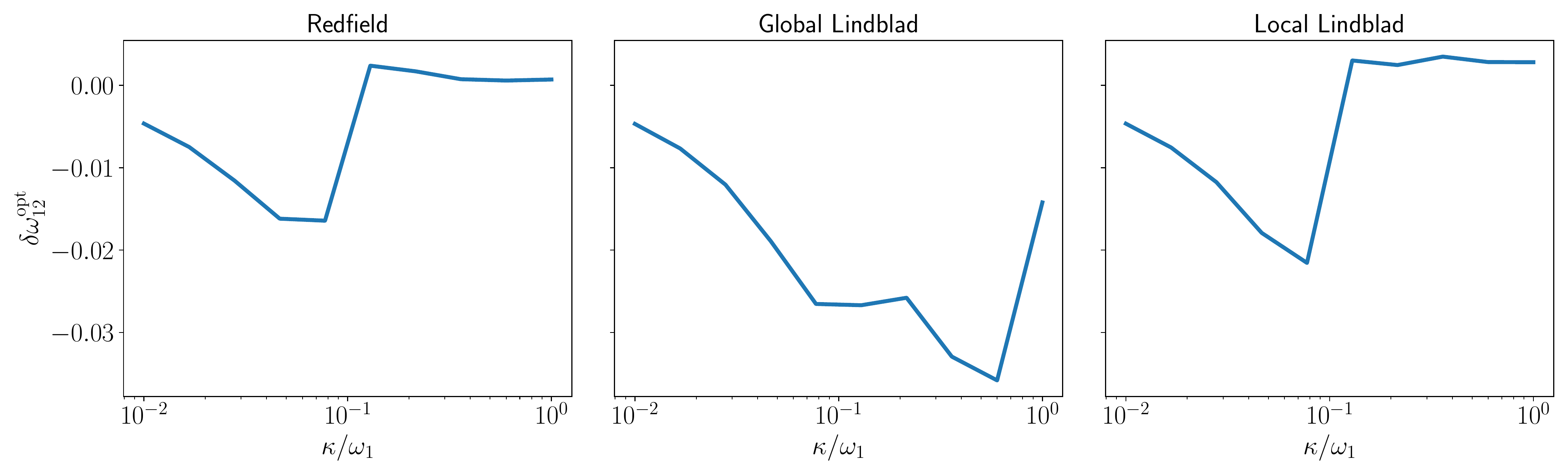}
\caption{Optimized values of $\delta \omega_{12}^\mathrm{opt}$ as a function of non-optimized value of $\kappa$ at low bare temperature $\hbar \beta \omega_1 = 5$, $g = 0.1 \omega_1$. The optimization is done by minimizing the  value of $\Delta_\mathrm{max}$ as a function of $\kappa$, $\beta$, $a_1$, $a_2$, and $a_3$.}
\end{figure*}

\begin{figure*}[!ht]
\includegraphics[width=\linewidth]{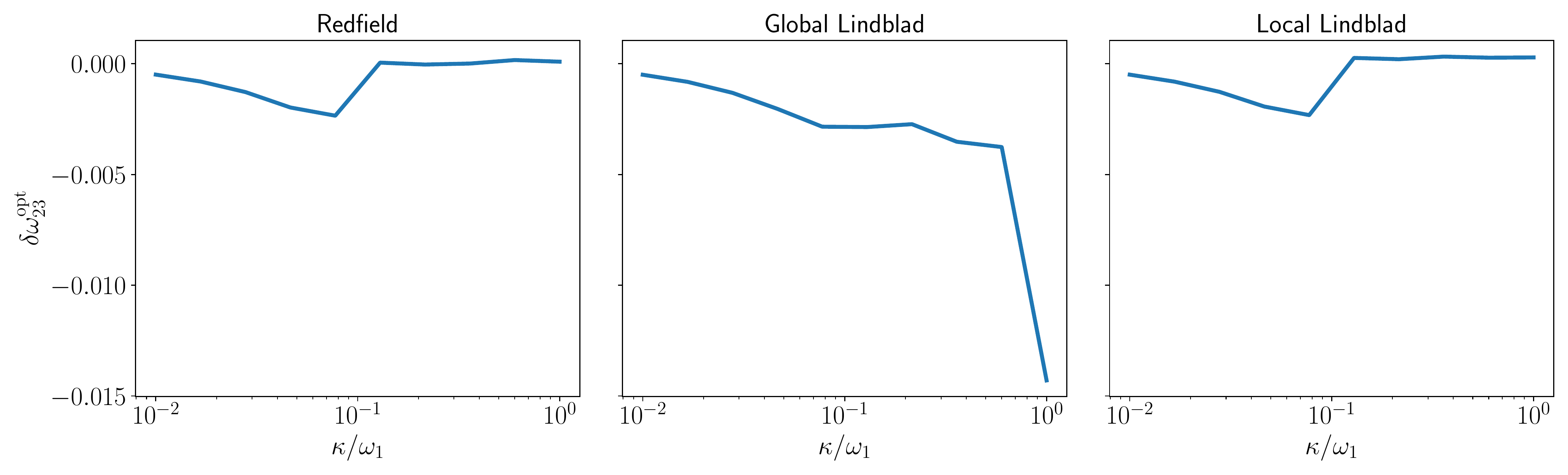}
\caption{Optimized values of $\delta \omega_{23}^\mathrm{opt}$ as a function of non-optimized value of $\kappa$ at low bare temperature $\hbar \beta \omega_1 = 5$, $g = 0.1 \omega_1$. The optimization is done by minimizing the  value of $\Delta_\mathrm{max}$ as a function of $\kappa$, $\beta$, $a_1$, $a_2$, and $a_3$.}
\end{figure*}

\begin{figure*}[!ht]
\includegraphics[width=\linewidth]{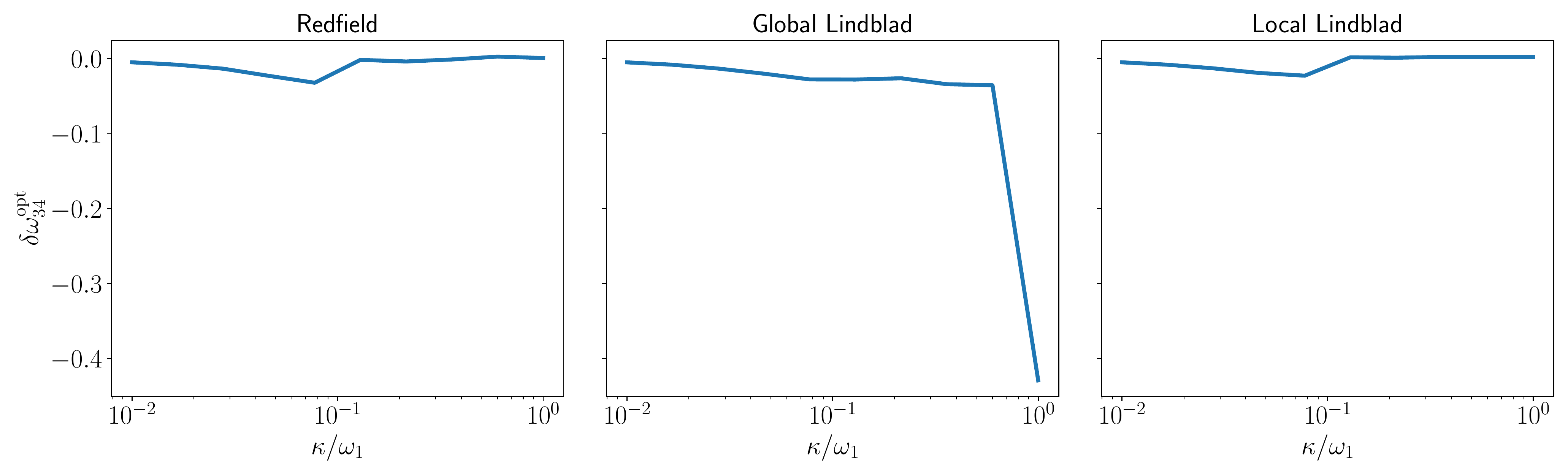}
\caption{Optimized values of $\delta \omega_{34}^\mathrm{opt}$ as a function of non-optimized value of $\kappa$ at low bare temperature $\hbar \beta \omega_1 = 5$, $g = 0.1 \omega_1$. The optimization is done by minimizing the  value of $\Delta_\mathrm{max}$ as a function of $\kappa$, $\beta$, $a_1$, $a_2$, and $a_3$.}
\end{figure*}

\acknowledgments

This research  was  financially  supported  by  the  European  Re\-search  Council  under  Grant  No.~681311  (QUESS), by the Academy of Finland through its Centre of Excellence in Quantum  Technology  (QTF)  (Grant  Nos.~312298  and~312300), by the Jane and Aatos Erkko Foundation, and
by the Technology Industries of Finland Centennial Foundation. It was also supported by the German Science Foundation (Grant Nos. AN336/11-1 and AN336/12-1), the Centre for Integrated Quantum Science and Technology (IQ$^{\rm ST}$), and the Zeiss Foundation under the Grant TQuant. The authors wish to acknowledge CSC~-- IT Center for Science, Finland, for computational resources.